\documentclass{article}
\usepackage{arxiv}
\usepackage{graphicx} 
\newif\ifclean
\cleantrue 
\usepackage{graphicx}
\usepackage{cite}
\usepackage{amsmath}
\usepackage{amssymb}
\usepackage{subcaption}
\usepackage{algorithm}
\usepackage{algpseudocode}
\usepackage{amsthm}
\usepackage{framed}
\newtheorem*{remark}{Remark}
\usepackage{wrapfig}
\usepackage{hyperref}
\usepackage{listings}
\usepackage[perpage]{footmisc}

\ifclean
\newcommand{\ADD}[1]{{#1}}
\newcommand{\COMMENT}[1]{{}}
\newcommand{\QUESTION}[1]{{}}
\newcommand{\TODO}[1]{{}}
\else
\usepackage[dvipsnames]{xcolor}
\usepackage{soul}
\newcommand{\COMMENT}[1]{\textcolor{cyan}{{[ \sc{#1} ]}}\\} 
\newcommand{\QUESTION}[1]{\textcolor{blue}{{[{QUESTION:} \it{#1} ]}}\\} 
\newcommand{\TODO}[1]{\textcolor{red}{TODO: {\bf{#1}}}\\} 

\newcommand{\red}[1]{\textcolor{red}{{#1}}}
\newcommand{\caution}{\red{\bf Draft: \today. Do not distribute.}}
\newcommand{\ADD}[1]{\textcolor{blue}{{{#1}}}}
\fi

\usepackage{upgreek}

\newcommand{\aref}[1]{App.\,\ref{#1}}
\newcommand{\fref}[1]{Fig.\,\ref{#1}}
\newcommand{\tref}[1]{Table\,\ref{#1}}
\newcommand{\eref}[1]{Eq.\,(\ref{#1})}

\newcommand{\sref}[1]{Sec.\!~\ref{#1}}

\newcommand{\cref}[1]{Ref.\,\cite{#1}}

\newcommand{\eb}{\mathbf{e}}

\newcommand{\gb}{\mathbf{g}}

\newcommand{\hb}{\mathbf{h}}

\newcommand{\kb}{\mathbf{k}}
\newcommand{\xb}{\mathbf{x}}
\newcommand{\zb}{\mathbf{z}}

\newcommand{\yb}{\mathbf{y}}
\newcommand{\Ab}{\mathbf{A}}
\newcommand{\Bb}{\mathbf{B}}
\newcommand{\Cb}{\mathbf{C}}

\newcommand{\Eb}{\mathbf{E}}

\newcommand{\Pb}{\mathbf{P}}

\newcommand{\Lb}{\mathbf{L}}
\newcommand{\Mb}{\mathbf{M}}
\newcommand{\Ib}{\mathbf{I}}
\newcommand{\Rb}{\mathbf{R}}

\newcommand{\Xb}{\mathbf{X}}
\newcommand{\Yb}{\mathbf{Y}}
\newcommand{\Zb}{\mathbf{Z}}
\newcommand{\Ac}{\mathcal{A}}
\newcommand{\Bc}{\mathcal{B}}

\newcommand{\Ic}{\mathcal{I}}

\newcommand{\Zc}{\mathcal{Z}}

\newcommand{\chib}{{\boldsymbol{\chi}}}

\newcommand{\Omegab}{{\boldsymbol{\Omega}}}

\newcommand{\tr}{{\operatorname{tr}}}

\newcommand{\partialb}{\boldsymbol{\partial}}

\newcommand{\NN}{\mathsf{N}\!\mathsf{N}}

\usepackage[T1]{fontenc}

\newcommand{\density}{\rho}

\newcommand{\defgrad}{\mathbf{F}}

\newcommand{\abstemperature}{T} 

\newcommand{\heatflux}{\mathbf{q}}

\newcommand{\pkstress}{\mathbf{S}}

\newcommand{\conjugateforce}{{\mathbf{k}}}

\newcommand{\lagrangestrain}{\mathbf{E}}

\newcommand{\internalenergy}{\epsilon}

\newcommand{\internalvariables}{{\boldsymbol{\kappa}}}

\newcommand{\entropy}{\eta}
\newcommand{\freeenergy}{\psi}
\newcommand{\heatgeneration}{R}

\newcommand{\dissipation}{\Gamma}

\newcommand{\dissipationpotential}{\phi}
\newcommand{\bipotential}{\beta}

\newcommand{\Grad}{\boldsymbol{\nabla}_{\mathbf{X}}}

\newcommand{\zerob}{\mathbf{0}}

\newcommand{\Sup}{\operatorname{sup}}
\newcommand{\flow}{\mathbf{r}}
\graphicspath{{{./}}}
\title{A Hierarchy of Thermodynamic Learning Frameworks for Inelastic Constitutive Modeling}
\renewcommand{\shorttitle}{\bf Learning frameworks for modeling inelasticity}
\author{
Reese E. Jones\\
Sandia National Laboratories \\
Livermore CA, USA
\And
Jan N. Fuhg\footnotemark[1] \\
The University of Texas at Austin \\
Austin TX, USA
}

\date{}

\begin{document}
\setlength\parindent{10pt}
\maketitle

\footnotetext[1]{Corresponding author: \texttt{jan.fuhg@utexas.edu}}

\begin{abstract}
Recent advances in physics-augmented neural networks have enabled thermodynamically consistent, data-driven constitutive modeling of complex, inelastic materials.
Most existing approaches, however, implicitly adopt a specific thermodynamic framework and embed structural assumptions such as flow normality, dual dissipation potentials, or other structure from classical, manually-constructed models directly into the learning architecture.
Consequently, differences in predictive performance may arise not only from data or network design, but also from the underlying theoretical assumptions.
In this work, we present a unified comparison of several thermodynamically consistent inelastic modeling frameworks in a machine learning context.
We consider internal-variable formulations with dissipation potential, generalized standard materials, and metriplectic structures.
We analyze their structural assumptions, admissible dependencies, convexity requirements, and implications for dissipation and evolution.
Each framework is implemented within a common neural potential architecture based on invariant representations and neural ordinary differential equations.
This unified setting \ADD{reduces architectural confounding and allows us to assess, to the extent possible, how differences in thermodynamic structure affect predictive performance.}
The models are trained and evaluated on three representative inelastic datasets generated from high-fidelity representative volume element simulations: an elastoplastic alloy, a viscoelastic composite, and a rate-dependent crystal plasticity polycrystal.
By isolating the role of thermodynamic structure, we assess how restrictions such as duality, normality, operator-based evolution, and convexity influence learnability, expressiveness, stability, and generalization.
The results show that all principled frameworks can produce accurate predictions on held-out data.
The most structurally constrained showed some marginal deficiencies on the most complex dataset (an elastic-plastic microstructural sample) but had a marginal advantage on a corresponding homogeneous material dataset.
\end{abstract}


\raggedbottom

\section{Introduction}

Recently, the art of constitutive modeling has experienced rapid growth due to the use of machine learning techniques.
While manually constructed phenomenological models have been successful in many applications, systematic model selection \cite{linhart1986model,ding2018model} is rarely applied; instead, they rely on preselected functional forms and evolution equations, which can lead to significant discrepancies and make systematic model construction, calibration, and automation cumbersome for complex materials.
On the other hand, the success of machine learning-based constitutive models relies on more general functional representations and the deliberate introduction of physically motivated inductive learning biases, which restrict the hypothesis space in accordance with known mechanical and thermodynamic principles \cite{fuhg2025review}.
Both  \emph{physics-augmented neural networks} (PANNs)~\cite{klein2022finite,rosenkranz2024viscoelasticty,fuhg2024extreme,benady2024nn,jailin2024experimental,jadoon2025inverse,tepole2025polyconvex,kalina2025physics,zlatic2024recovering}, and related approaches such as \emph{constitutive artificial neural networks} (CANNs)~\cite{linka2021constitutive,linka2023new,holthusen2024theory,boes2024accounting,holthusen2024polyconvex,abdolazizi2024viscoelastic,thakolkaran2025can}, adhere to this approach.
Likewise, structure-preserving learning frameworks and similar concepts \cite{greydanus2019hamiltonian,cranmer2020lagrangian,chen2019symplectic,gonzalez2019thermodynamically,gruber2025efficiently} have also been applied to physical contexts outside continuum mechanics.

In fact, data-driven constitutive modeling approaches, like PANNs, implicitly commit to a particular thermodynamic structure.
Choices such as the existence of an underlying free energy, the form of dissipation, the role of internal variables, or the assumption of associative flow are embedded directly into the model architecture or loss function, based on \textit{a priori} beliefs about the nature of the data and its compatibility to classical theories of continuum mechanics and thermodynamics.
As a result, different learning frameworks may encode fundamentally different physical assumptions, yet are rarely compared on equal footing in terms of their theoretical implications or their impact on data-driven model performance.
Across many formulations, a unifying assumption is the existence of an energy potential, usually the Helmholtz free energy, that governs the reversible response of the material.
The primary distinctions between frameworks then emerge in how irreversibility is modeled, namely through the introduction of internal variables, evolution equations, dissipation potentials, or more general non-equilibrium structures, all of which are subject to the constraints imposed by the second law of thermodynamics.

Within this thermodynamic setting, several distinct yet related theoretical frameworks have been developed to model inelastic material behavior.
These frameworks share foundational principles, yet differ in how state variables are introduced, how dissipation is structured, and how evolution equations are generated.
Some of the commonly employed formulations can be summarized as:
\begin{itemize}
\item \textbf{Internal state variable (ISV) theories:} Inelastic behavior is represented through a finite set of internal variables whose evolution is modeled by an ordinary differential equation, with stresses derived from a free energy. The evolution is constrained only by the second law \cite{coleman1967thermodynamics}.
\item \textbf{ISV theories with a dissipation potential (DP):} Evolution of internal variables is governed by a dissipation potential, providing a variational structure and enforcing dissipation through convexity of this potential, as in Chaboche-type models \cite{lemaitre1994mechanics}.

\item \textbf{Generalized Standard Materials (GSM):} Reversible and irreversible responses are derived from thermodynamic potentials linked by Legendre–Fenchel duality, leading to normality conditions and an Onsager-like thermodynamic structure \cite{halphen1975materiaux}.
\item \textbf{Metriplectic (MP) / General Equation for Non-Equilibrium Reversible-Irreversible Coupling (GENERIC) formulations:} Non-equilibrium dynamics are described by the superposition of Hamiltonian (energy-conserving) and dissipative (entropy-producing) flows, offering a geometric and operator-based approach to thermodynamic evolution \cite{morrison1984bracket,ottinger1997dynamics}.
\end{itemize}
While these frameworks \ADD{have} been extensively studied, their adaptation to data-driven constitutive modeling is relatively nascent.

Existing machine-learning-based approaches typically focus on a single formulation, embedding its assumptions as fixed inductive biases without contrasting the chosen framework against alternatives.
Consequently, there is a lack of a unified, machine-learning-focused comparison that clarifies how different theoretical frameworks influence learnability, expressiveness, generalization, and physical consistency when models are trained from data.
This gap makes it difficult to assess whether observed performance differences arise from the data, the learning architecture, or the underlying thermodynamic assumptions themselves.

In particular, the following key questions arise:
\begin{itemize}
\item What structural assumptions does each framework impose on constitutive models?
\item What restrictions do these assumptions introduce in the expressiveness of a data-driven model?
\item How do these choices affect generalization, interpretability, and physical consistency of the predictions?
\end{itemize}
In this work, we address these questions by combining a concise theoretical review of inelastic constitutive frameworks with a data-driven perspective.
We emphasize the structural assumptions that are most relevant for machine learning, rather than providing an exhaustive historical account.
Building on this foundation, we implement each framework within a unified computational setting based on neural representations of thermodynamic potentials, enabling a controlled and consistent comparison.
The resulting models are then evaluated side-by-side on a variety of representative inelastic material datasets without closed-form representations \ADD{to clarify the influence of the underlying thermodynamic structure while keeping the neural implementation and training procedure as consistent as possible.}

In the next section, \sref{sec:background}, we review the thermodynamic foundations of inelastic constitutive modeling and introduce the theoretical frameworks considered in this work from a data-driven perspective.
\sref{sec:theory} presents the mathematical formulation of each framework and highlights its structural similarities and differences with the other frameworks.
\sref{sec:implementation} describes the unified neural implementation and training strategy.
\sref{sec:data} describes the datasets used for comparison.
The results and comparative analysis are presented in \sref{sec:results} and discussed in \sref{sec:discussion}.
Finally, \sref{sec:conclusion} summarizes the main findings and discusses implications for future data-driven constitutive modeling efforts.

\section{Background} \label{sec:background}

Data-driven constitutive modeling \ADD{can} operate in challenging regimes where data are sparse, heterogeneous, and expensive to obtain, particularly for complex inelastic phenomena involving history dependence and evolving microstructure.
In such settings, purely data-driven approaches are prone to nonphysical extrapolation and poor generalization unless augmented with strong architectural priors \cite{fuhg2024review}.
In classical constitutive modeling, these prior conceptions arise naturally through thermodynamic principles, which restrict the admissible form of material response by enforcing energy conservation, dissipation, objectivity, etc.
Physics-augmented machine-learning-based constitutive models implicitly inherit these same restrictions through architectural choices, loss functions, and/or training constraints.
As a result, different learning frameworks correspond not merely to different regression strategies, but to different choices of theoretical structure, in the sense of \emph{model selection} \cite{linhart1986model,ding2018model}, beyond the architectural details of the learning components.
Understanding how these structures arise in classical theory is therefore essential for interpreting, comparing, and designing data-driven, physics-augmented constitutive models.

\paragraph{Thermodynamic theory}
Classical theories of inelasticity formalize these structural restrictions primarily through the distinction between conserved and dissipative aspects, and equilibrium and non-equilibrium thermodynamics.

Equilibrium thermodynamics emerged from the foundational works of Carnot, Joule, and Clausius~\cite{carnot1824reflections,joule1843xxxii,clausius1850ueber}, and describes idealized states in which entropy production vanishes, and processes occur infinitely slowly.
In this setting, material behavior can be characterized entirely by \emph{state functions}, such as stress, temperature, and entropy, as formalized through the \emph{thermodynamic potentials} introduced by Helmholtz, Thomson, Maxwell, and Gibbs~\cite{von1847ueber,Thomson_1853,clerk1875xxii,Gibbs1873Graphical}.

Inelastic deformation, by contrast, involves irreversible processes that occur at finite rates and therefore lie outside equilibrium, requiring a description of how the system evolves subject to the second law.
Non-equilibrium thermodynamics \cite{onsager1931reciprocalI,onsager1931reciprocalII,prigogine1963introduction,de2013non,lebon2008understanding} provides this extension by introducing rates, fluxes, and additional variables to represent irreversible processes and history dependence, but much of this work focused on fluids and chemistry, not solids.
Early thermodynamic treatments of irreversibility by Duhem and Bridgman \cite{duhem2015ii,bridgman1923thermal} established dissipation and entropy production as fundamental constraints on admissible processes, while emphasizing their physical and mechanical interpretation in real materials (for a historical review, refer to \cref{horstemeyer2010historical}).
A significant conceptual step was taken by Onsager~\cite{onsager1931reciprocalI,onsager1931reciprocalII}, who introduced additional, non-observable variables to account for irreversible processes near equilibrium.
These variables, which are assumed \ADD{to} represent (microscopic/statistical) internal degrees of freedom governing dissipation, provide the first appearance of what are now called \emph{internal state} variables.
Furthermore, Onsager showed that, in the near-equilibrium regime, linear relations between thermodynamic forces and fluxes satisfy the second law if and only if the associated phenomenological coefficients of the fluxes obey reciprocal symmetry relations.
This result implies the existence of a quadratic dissipation function and an equivalent variational principle, establishing a prototype for dissipation-potential-based evolution laws.
Eckart~\cite{eckart1940thermodynamics} subsequently embedded Onsager’s force--flux framework into continuum mechanics by formulating local balance laws for entropy production that are compatible with stress, strain rate, and heat flux.
This step translated Onsager’s internal variables from an abstract thermodynamic setting into a form applicable to deformable continua and laid the groundwork for internal-variable formulations in solid mechanics.
Related mechanical interpretations of internal variables were further developed in the context of microstructure by Kr\"{o}ner \cite{kroner1965internal}.
The internal state variable concept was formalized as a systematic constitutive framework by Coleman and Gurtin~\cite{coleman1967thermodynamics}, who postulated that the thermodynamic state of an inelastically deforming material is determined by the observable variables together with a finite set of internal state variables.%
\footnote{
Later, Maugin \cite{maugin1994thermodynamics} introduced the idea of internal variables relaxing to equilibrium, and Reese and Govindjee \cite{reese1998theory} developed a nonlinear Maxwell model with equilibrium and non-equilibrium potentials.}
While the Coleman--Gurtin framework establishes a thermodynamically consistent state space, it leaves the evolution equations for the internal variables largely unspecified beyond the requirement of non-negative dissipation.
The \emph{local equilibrium} hypothesis, i.e., equilibrium thermodynamic relationships hold in local environments in real processes, is typically assumed.
Subsequent developments can be interpreted as different strategies for closing this constitutive gap.

An early systematic closure was provided by \emph{generalized standard materials} (GSM), a theory devised by Halphen and Nguyen~\cite{halphen1975materiaux,germain1983continuum}.
In GSM, the evolution of internal variables is derived from a dissipation potential dual to the free energy through convex analysis and variational principles, leading to normality and associated flow.
This additional structure guarantees thermodynamic consistency by construction, but significantly restricts the class of admissible constitutive behaviors.
Chaboche \cite{lemaitre1994mechanics} subsequently developed a widely-used class of internal-variable constitutive models that retain the thermodynamic framework of Coleman and Gurtin while relaxing the strict duality and normality requirements of GSM.
By allowing the yield surface, flow rule, and internal variable evolution equations to be specified independently, Chaboche-type models accommodate non-associated flow, complex cyclic behavior, and multiple hardening mechanisms, and have proven particularly effective in engineering applications.
Alternative mathematical frameworks have also been explored, although more thoroughly in fluid mechanics.
Metriplectic \cite{morrison1984bracket} (as in \emph{metric} and \emph{symplectic}) and GENERIC \cite{ottinger1997dynamics} formulations were developed for system-level thermodynamics; they can be adapted to the evolution of internal variables and impose structural constraints on admissible internal dynamics.
These essentially identical frameworks restrict the admissible form of the dynamics by decomposing it into reversible and irreversible components governed by distinct symplectic and metric operators.
More discussion of alternative modeling frameworks will be given in \sref{sec:theory}.

\paragraph{Physics-augmented data-driven models}
Motivated by these theoretical foundations, a rapidly growing number of developments have used machine learning for constitutive modeling of inelastic materials.
Rather than treating constitutive response as an unconstrained regression problem, many successful approaches embed physical assumptions into the learning architecture, such as energy conservation/dissipation and objectivity.
We observe that learning frameworks in which thermodynamic admissibility and satisfaction of other principles are guaranteed by the structure of the model class itself, rather than penalized or learned from data, are advantageous.

An early adoption of the internal state variable (ISV) framework in a data-driven setting \cite{jones2022neural} utilized a neural ODE-based model \cite{chen2018neural,dupont2019augmented}; it targeted inelastic homogeneous and microstructured materials and was shown to be a capable representation of elastic-plastic and viscoelastic behavior.
To the authors’ knowledge, the approach of Flaschel et al.~\cite{flaschel2023automated} is the first data-driven framework that explicitly adopts the generalized standard materials (GSM) structure, including a dual dissipation potential and variationally derived evolution equations.
Subsequent neural-network-based formulations embed this explicit GSM structure by parameterizing the free energy and the dual dissipation potential with convex or monotonic neural networks, enabling thermodynamic consistency by construction while learning constitutive behavior from data~\cite{rosenkranz2024viscoelasticty,flaschel2025convex,kalina2025physics}.
By contrast, a number of recent data-driven constitutive modeling approaches combine neural networks with thermodynamic structure without explicitly adopting the dual GSM formulation.
In these methods, neural networks are used to parameterize a Helmholtz free energy together with a dissipation-related potential, while architectural constraints such as convexity, monotonicity, or invariance are imposed to ensure objectivity and positive dissipation by construction~\cite{tacc2023data,holthusen2024polyconvex,jones2025physics,holthusen2026complement}.
Inelastic evolution is typically formulated through stress- or force-driven flow rules enabling the learning of history-dependent behavior from stress--strain data alone.

\begin{wraptable}{r}{0.52\textwidth}
\vspace{-10pt}
\centering
\begin{tabular}{|l|c|}
\hline
\textbf{Name} & \textbf{Symbol} \\
\hline
\hline
Internal energy & $\internalenergy$ \\
(Absolute) temperature  & $\abstemperature$ \\
(Helmholtz) free energy & $\freeenergy = \internalenergy - \abstemperature \entropy$ \\
Entropy                 & $\entropy = -\partialb_\abstemperature \freeenergy$ \\
Dissipation potential   & $\dissipationpotential$ \\
Dissipation             & $\dissipation$ \\
\hline
Time                    & $t$ \\
Reference position      & $\Xb$ \\
Current position        & $\xb$ \\
Motion & $\xb = \chib(\Xb,t)$ \\
Deformation gradient    & $\defgrad = \partialb_\Xb \chib$ \\
(Right Cauchy-Green) \ADD{deformation} tensor & $\Cb = \defgrad^T \defgrad$ \\
(Lagrange) strain tensor & $\lagrangestrain = \tfrac{1}{2}(\Cb-\Ib)$ \\
\hline
(Second Piola-Kirchhoff) stress tensor$^*$ & $\pkstress = \partialb_\lagrangestrain \freeenergy \ADD{ = 2 \partialb_\Cb \freeenergy}$ \\
(Reference) heat flux vector & $\heatflux$ \\
\ADD{(Reference) heat generation} & $R$ \\
\hline
Internal state   &  $\internalvariables$ \\
Conjugate force  &  $\conjugateforce = \ADD{-}\partialb_{\internalvariables} \freeenergy$ \\
Flow             &  $\flow = \dot{\internalvariables}$ \\
\hline
\end{tabular}
\caption{Notation.
\ADD{$^*$Note this is only part of the stress definition for GSM, where rates also contribute to the stress/momentum flux.}}
\label{tab:notation}
\vspace{-10pt}
\end{wraptable}

Although some of these approaches are motivated by the GSM framework and share its use of energetic and dissipative potentials, the evolution equations do not arise from an explicit primal-dual dissipation structure based on Legendre-Fenchel duality.
Hence, they are structurally closer to Chaboche-type internal-variable models.
In contrast, most recent data-driven efforts adopting metriplectic or GENERIC structure focus on learning thermodynamic generators or operators at the level of the global dynamical system, using regression, neural networks, or physics-informed architectures, with primary applications in dynamical system identification, reduced-order modeling, or surrogate simulation \cite{gonzalez2019thermodynamically,simavilla2024hammering,hernandez2023port}.
Direct application of metriplectic or GENERIC formulations to constitutive modeling remains comparatively rare; although, a number of authors \cite{mielke2011formulation,ghnatios2019data,schiebl2021structure}, most notably Hutter and co-workers \cite{hutter2011formulation,hutter2008continuum,hutter2008finite,hutter2015statistical}, have applied the framework to constitutive modeling of inelastic materials.
Nevertheless, most of the resulting formulations are problem-specific and generally do not constitute systematic, extensible constitutive modeling frameworks comparable to those developed within GSM or ISV-based approaches.

\section{Theory} \label{sec:theory}

Although elasticity theory is the most mature and agreed-upon aspect of constitutive modeling, many technologically relevant processes are dissipative and inelastic.
There are a number of widely accepted principles that guide inelasticity theory.
These, together with some general considerations, form the foundation of inelasticity theory.
Here we outline four inelastic modeling frameworks, starting with \emph{internal state variable} theory \cite{coleman1967thermodynamics}, with the goal of highlighting their similarity and differences.
Loosely speaking, \ADD{some theories} can be seen as a restriction of \ADD{others} via more structure and, hence, embedded phenomenology.
Here we use notation summarized in \tref{tab:notation} and consider the non-isothermal context, but do not focus on heat flow.
\ADD{
In a non-equilibrium scenario, the temperature gradient may also play a role but can be eliminated as a dependency in some response functions \cite{coleman1967thermodynamics}.
Here we make the common simplifying assumption to omit dependence on temperature gradients in the mechanical response functions.
}

\subsection{Principles}

A starting point for rational mechanics treatments of inelasticity is a Helmholtz free energy that gives a sequence of elastic materials indexed by internal \emph{state} variables \cite{prigogine1963introduction,coleman1967thermodynamics}.
Other frameworks take a similar sequence-of-models approach \cite{grmela2010generic} to represent the evolution of a dissipative system.
Unlike stress, deformation, and temperature, these internal variables are largely unobservable and hence their number and definition are subject to debate.
This state-based treatment of history/path dependence contrasts with memory kernel treatments that are generally less adaptable.%
\footnote{Note the change of approach from a memory kernel to a finite-dimensional state is warranted by the assumption of \emph{fading memory} \cite{wang1965principle,truesdell2004non}.}

The thermodynamic state and potentials that modeling frameworks are based on are not unique.
The Legendre-Fenchel (LF) transform:
\begin{equation}
\phi^*(\Yb, \Zb) = \Sup_{\Xb} \bigl(  \phi(\Xb, \Zb) - \Xb \cdot \Yb \bigr) \ ,
\end{equation}
provides one route to change the potential and its dependent variables through the property $\Yb = \partialb_\Xb \phi$.
For instance, internal energy $\internalenergy$ in terms of the entropy $\entropy$ results from the transformation of the Helmholtz free energy $\freeenergy$:
\begin{equation} \label{eq:helmholtz}
\freeenergy(\lagrangestrain, \abstemperature) =
\Sup_{\entropy}  \internalenergy(\lagrangestrain,     \entropy) - \entropy \abstemperature \ ,
\end{equation}
with respect to the absolute temperature $\abstemperature$.
The potential, its primary kinematic variable, and the conjugate flux form a generalized work-conjugate triple.
We will focus on the Helmholtz free energy $\freeenergy$, Lagrange strain $\lagrangestrain$, and second Piola-Kirchhoff stress $\pkstress = \partialb_{\lagrangestrain} \freeenergy$; however, we will not concentrate on kinematics other than staying in the finite deformation context.

Materials with such a thermodynamic potential are subject to the Euler balances and the second law of thermodynamics.
The local balance of energy is:
\begin{equation} \label{eq:bal_eng}
\density_0 \dot{\internalenergy} = \pkstress \cdot \dot{\lagrangestrain} - \Grad \cdot \heatflux + \density_0 \ADD{\heatgeneration}  \ ,
\end{equation}
Colloquially, the second law of thermodynamics requires that the entropy, as a measure of disorder, does not decrease in a closed system.
More formally, the Clausius-Duhem statement of the second law in local form:
\begin{equation} \label{eq:CD}
\density_0 \dot{\entropy} - \density_0 \frac{\ADD{\heatgeneration}}{\abstemperature} - \Grad \cdot \frac{\heatflux}{\abstemperature}
\ge 0 \ ,
\end{equation}
plays a central role in constraining mechanics models.
This reduces to
\begin{equation} \label{eq:dissipation}
\dissipation \equiv \pkstress \cdot \dot{\lagrangestrain} - \dot{\freeenergy}
\ge 0 \ ,
\end{equation}
assuming the Fourier inequality $ \heatflux \cdot \Grad \abstemperature \le 0$ holds and the system is isothermal $\dot{\abstemperature} = 0$.

The Coleman-Noll (C-N) procedure \cite{coleman1974thermodynamics}, starting with the Clausius-Duhem inequality \eqref{eq:CD} results in:
\begin{equation} \label{eq:C-N}
(\pkstress - \partialb_{\lagrangestrain} \freeenergy) \cdot \dot{\lagrangestrain}   +
(\entropy + \partial_\abstemperature \freeenergy) \dot{\abstemperature} +
(  - \partialb_{\internalvariables} \freeenergy) \cdot \dot{\internalvariables}
-  \frac{1}{\density_0 \abstemperature} \heatflux \cdot \Grad \abstemperature
\ge 0
\end{equation}
where the rates are assumed to act like arbitrary variations so that the (\emph{local equilibrium}) thermodynamic connections:
\begin{eqnarray}
\pkstress       &=& \partialb_{\lagrangestrain} \freeenergy \\
\entropy        &=& -\partial_\abstemperature \freeenergy
\end{eqnarray}
can be made.
This is inherently a local-equilibrium argument.
Here $\internalvariables$ represents any additional dependencies of the free energy $\freeenergy$, such as internal states.
With the Fourier inequality and these identities, \eref{eq:dissipation} becomes:
\begin{equation}
\dissipation = - \partialb_{\internalvariables} \freeenergy \, \cdot \dot{\internalvariables}  \ge 0 \, .
\end{equation}

Additional considerations, such as invariance/equivariance of inputs and outputs with respect to change of coordinates/material frame, and material stability/polyconvexity, are important and will be addressed in \sref{sec:implementation}, but we will not discuss them in detail here; please refer to \cref{fuhg2024polyconvex} for theoretical developments, for instance.

\subsection{Assumptions}

Most applications of constitutive modeling are implicitly based on the axioms of a \emph{simple} material \cite{noll1972new,truesdell2004non, smith1993introduction}, such as: (a) local dependence, i.e. no higher order gradients than the first order deformation gradient, and (b) dependence only on the history of the deformation gradient (and temperature) at a material point.
There are important applications where stochastic or representative volume samples of the material microstructure or other non-locality need to be treated, and some of the simple material results \cite{cimmelli2010generalized} have been extended to these cases; however, we will focus on simple (homogeneous) materials.

Another fundamental consideration is that \emph{state} functions, such as temperature and stress, are generally only well defined in thermodynamic equilibrium.
This generally requires that the relevant relaxation processes occur on timescales that are short and length-scales that are long with respect to the time- and length-scales of observation.
The dimensionless \emph{Deborah} number in rheology quantifies this notion \cite{maugin1994thermodynamics}.
Different quantities can be connected to different relaxation mechanisms (such as dislocations for deformation and electrons for heat transfer in metals) and hence scales.
Reversibility also plays a role, for example, the elastic deformation of a perfect single crystal and an amorphous glass are both usually considered steady states, although the relaxation times tend to zero and infinity, respectively.
Typically, \emph{local equilibrium} is assumed \cite{maugin1994thermodynamics}, whereby the equilibrium relations such as $\pkstress = \partialb_\lagrangestrain \freeenergy$, hold in potentially rapid and heterogeneous processes.
Other approaches, such as extended irreversible thermodynamics \cite{jou1988extended,jou1999extended} where fluxes are treated as independent fields and temperature has a dependence on heat flux, exist but are not prevalent in continuum mechanics.

\subsection{Formulations}

In this section, we frame a number of currently used material modeling frameworks in terms of their structure, assumptions, and the dynamics of the hidden states.
We are not necessarily trying to establish a taxonomy, merely contrasting the frameworks in light of their utility in machine learning.
The formulations are introduced \ADD{largely} in order of increasing structural restriction, beginning with the least constrained closure and progressing toward frameworks that impose progressively stronger thermodynamic structure, \ADD{although the last formulation is capable of representing the first three via specific choices}.
\tref{tab:formulation_distinctions} summarizes distinctions between the formulations under consideration.

\paragraph{Internal state variable (ISV)}
Coleman and Gurtin (C-G) \cite{coleman1967thermodynamics} proposed a general theory of inelasticity based on internal state variables (ISVs), which was later expanded upon by numerous authors \cite{kratochvil1969thermodynamics,maugin1994thermodynamics,mcdowell2005internal}.
Following standard (equilibrium) thermodynamic theory,  stress is defined as the derivative of the Helmholtz free energy with respect to the deformation measure:
\begin{equation} \label{eq:stress}
\pkstress = \partialb_{\lagrangestrain} \freeenergy(\lagrangestrain, \internalvariables, \abstemperature) \ ,
\end{equation}
while the flow of internal states $\internalvariables$ is given by an ordinary differential equation (ODE):
\begin{equation}
\dot{\internalvariables} = \flow(\lagrangestrain, \internalvariables, \abstemperature) \ ,
\end{equation}
where inputs are given by the \emph{equipresence} assumption \cite{truesdell2004non}, but are generally unrestricted.
C-G do not prescribe structure to the flow $\flow$ other than the dissipation requirement:
\begin{equation} \label{eq:dissipation_req}
\dissipation  \equiv
\pkstress \cdot \dot{\lagrangestrain} - \dot{\freeenergy}
= -\partialb_{\internalvariables} \freeenergy \cdot \dot{\internalvariables}
= \conjugateforce \cdot \flow
\ge 0 \ ,
\end{equation}
resulting from the second law \eqref{eq:CD} and the definition of a conjugate force
\begin{equation} \label{eq:conjugate_force}
\conjugateforce \equiv -\partialb_{\internalvariables} \freeenergy  \ .
\end{equation}

Note there is no explicit dependence of the free energy on the deformation rate $\dot{\lagrangestrain}$ and no dissipation potential in the original formulation;%
\footnote{
Note Coleman-Gurtin \cite{coleman1967thermodynamics} rely on the Coleman-Noll argument \cite{coleman1974thermodynamics} to associate thermomechanical variables with potential derivatives. This argument relies on the rates being arbitrary and the functions multiplying the rates being independent of the rates, hence a free energy dependent on strain rate is incompatible with this formulation.
}
\ADD{hence, a general, potentially high-dimensional flow field $\flow$ needs to be modeled.}
Furthermore, yield-like behavior was not considered in the original ISV theory; however, it can be enhanced to accommodate switching behavior \cite{jones2025attention,jones2025physics}.

\paragraph{Dissipation potential (DP)}

After Coleman and Gurtin \cite{coleman1967thermodynamics}, Chaboche \cite{lemaitre1994mechanics} and others \cite{ziegler1963some,moreau1970lois} introduced a \emph{pseudopotential}  $\dissipationpotential$ to determine the flow:
\begin{equation} \label{eq:dp_update}
\dot{\internalvariables} = \partialb_\conjugateforce \dissipationpotential
\end{equation}
Imposing a convex (in $\conjugateforce$), non-negative dissipation potential that vanishes at the origin is \emph{a} means of enforcing the dissipation requirement \eqref{eq:dissipation_req}:
\begin{equation} \label{eq:dp_diss}
\dissipation = -\partialb_{{\internalvariables}} \freeenergy \cdot \dot{\internalvariables} =
\conjugateforce \cdot \partialb_\conjugateforce \dissipationpotential \ge 0 \ .
\end{equation}
There are other means, such as requiring a degree of homogeneity \cite{ottosen2005mechanics}
or monotonicity in positively homogeneous stress invariants \cite{holthusen2026complement}. However, the convexity of the dissipation potential remains the most common and widely adopted sufficient condition.
If a yield function is available, the dissipation potential is often formulated as the yield function plus a deviation, and hence, DP does not assume associative flow (with respect to yield).
Yield behavior and the rate-independent Karush-Kuhn-Tucker conditions complicate this formulation, as they do each of the frameworks we present.

This and the ISV framework are based on local-equilibrium assumptions; the next theory accommodates non-equilibrium thermomechanics.

\begin{remark}
As a segue to the next formulation,
the dissipation potential $\dissipationpotential$ has a dual dissipation potential $\dissipationpotential^*$ through the LF transform:
\begin{equation} \label{eq:dual_dissipation}
\dissipationpotential^*(\dot{\internalvariables}; \Zc) = \Sup_{\conjugateforce} \bigl(  \dot{\internalvariables} \cdot \conjugateforce - \dissipationpotential(\conjugateforce; \Zc) \bigr)
\end{equation}
such that  $\partialb_\Zc \dissipationpotential^* = -\partialb_\Zc \dissipationpotential$ where $\Zc=\{ \lagrangestrain, \dot{\lagrangestrain}, \internalvariables, \abstemperature, \ldots\}$ \ADD{are the variables auxiliary to the transform.}
With the dual potential, there is a connection between the free energy and dissipation potential via the conjugate force
\begin{equation} \label{eq:conj_force2}
\conjugateforce \equiv -\partialb_{\internalvariables} \freeenergy
=
\partialb_{\dot{\internalvariables}} \dissipationpotential^* \ ,
\end{equation}
where the first form is the definition \eqref{eq:conjugate_force}, and second is a direct result of the LF transform \eqref{eq:dual_dissipation}.
This is also known as the Biot relation and is difficult to enforce on the potentials $\freeenergy$ and $\dissipationpotential$ due to its differential nature.
Also, the LF transform \eqref{eq:dual_dissipation} leads directly to the identity
\begin{equation} \label{eq:dissipation_dual_identity}
\dissipation = \dissipationpotential + \dissipationpotential^* \ .
\end{equation}
\ADD{at the supremum.}
\end{remark}

\paragraph{Generalized standard material (GSM)}
As a departure from Coleman-Gurtin and related classical theory, \ADD{Halphen} and Nguyen \cite{halphen1975materiaux,nguyen1988mechanical} developed the \emph{Generalized} \emph{Standard} \emph{Material} (GSM) theory.
In GSM, fluxes are tied to potentials via a form of Onsager reciprocity \cite{onsager1931reciprocalI,onsager1931reciprocalII}.
Both the physical stress \cite{mcbride2018dissipation,flaschel2025convex}:
\begin{equation} \label{eq:gsm_stress}
\pkstress = \partialb_\lagrangestrain \freeenergy + \partialb_{\dot{\lagrangestrain}} \dissipationpotential^*
\end{equation}
and the internal forces:
\begin{equation}
\partialb_{\internalvariables} \freeenergy +  \partialb_{\dot{\internalvariables}} \dissipationpotential^*
\end{equation}
are additively split into equilibrium  and non-equilibrium contributions \cite{de2013non,lebon2008understanding} and connected to free energy
\begin{equation} \label{eq:gsm_free_energy}
\freeenergy = \freeenergy(\lagrangestrain,\internalvariables;\abstemperature)
\end{equation}
and dissipation potential
\begin{equation} \label{eq:dual_dissipation_potential}
\dissipationpotential^* = \dissipationpotential^*(\dot{\lagrangestrain},\dot{\internalvariables};\abstemperature) \ ,
\end{equation}
which are dependent on the observable deformation and internal state, and their rates, respectively.

Note the restriction of dependencies for free energy and dissipation potential,  $\freeenergy$ only depends on kinematic variables $\{\lagrangestrain,\internalvariables\}$,  and $\dissipationpotential^*$ only rates of these variables.%
\footnote{
Here, and in the other frameworks, we assume the temperature $\abstemperature$ acts as a variable parameter as opposed to a kinematic variable with a rate; it is also possible to have $\dissipationpotential^*$ depend on $\dot{\abstemperature}$ instead.}
Note this formulation invalidates the use of the Coleman-Noll rule \eqref{eq:C-N} to associate potential derivatives with fluxes due to the explicit dependence of the stress on the strain rate.
It is \ADD{postulated} that the \emph{Biot relation} \cite{biot1954theory,halphen1975materiaux}:
\begin{equation} \label{eq:biot}
\partialb_{\internalvariables} \freeenergy + \partialb_{\dot{\internalvariables}} \dissipationpotential^* \equiv \zerob \ ,
\end{equation}
which is interpreted as a microbalance \cite{stainier2013variational}), connects the free energy and dissipation potential.

With this formulation the (mechanical) dissipation requirement \eqref{eq:dissipation_req} reduces to:
\begin{equation} \label{eq:dissipation_gsm}
\dissipation
= ( \pkstress - \partialb_{\lagrangestrain} \freeenergy ) \cdot \dot{\lagrangestrain}  - \partialb_{\internalvariables} \freeenergy \cdot  \dot{\internalvariables}
=  \partialb_{\dot{\lagrangestrain}} \dissipationpotential^* \cdot \dot{\lagrangestrain}  + \partialb_{\dot{\internalvariables}} \dissipationpotential^* \cdot \dot{\internalvariables} \ge 0
\end{equation}
after using \ADD{the stress definition \eqref{eq:gsm_stress}} the Biot relationship \eqref{eq:biot}.
\eref{eq:dissipation_gsm} is an extended version of \eref{eq:dp_diss}, after noting the alternative definition of $\conjugateforce$ in \eref{eq:conj_force2}.
So it is sufficient for $\dissipationpotential^*(\dot{\lagrangestrain},\dot{\internalvariables};\abstemperature)$
to be convex, non-negative, and to attain its minimum at $(\dot{\lagrangestrain},\dot{\internalvariables})=(\boldsymbol{0},\boldsymbol{0})$
(e.g., $\dissipationpotential^*(\boldsymbol{0},\boldsymbol{0})=0$).

In this theory, the primal (rate-based) $\dissipationpotential^*$ and dual (force-based) $\dissipationpotential$ dissipation potentials must be \ADD{Legendre–Fenchel} duals of one another, i.e. :
\begin{equation} \label{eq:convex_conj}
\dissipationpotential(\conjugateforce;\Ac) = \sup_{\dot{\internalvariables}} \left( \conjugateforce \cdot \dot{\internalvariables} - \dissipationpotential^*(\dot{\internalvariables};\Ac) \right) ,
\end{equation}
and conversely,
\begin{equation} \label{eq:gsm_LF}
\dissipationpotential^*(\dot{\internalvariables};\Ac) = \sup_{\conjugateforce} \left( \conjugateforce \cdot \dot{\internalvariables} - \dissipationpotential(\conjugateforce;\Ac) \right) .
\end{equation}
\ADD{where the auxiliary variables $\Ac$ are subject to the ansatz \eqref{eq:dual_dissipation_potential}.}
In the primal (rate--based) formulation, Biot's relation \eqref{eq:biot} gives
\begin{equation} \label{eq:GSM_Biot_revised}
\conjugateforce = \partialb_{\dot{\internalvariables}} \dissipationpotential^*(\dot{\internalvariables};\dot{\lagrangestrain},\abstemperature)
\end{equation}
which (\emph{implicitly}) determines $\dot{\internalvariables}$.
\ADD{Note this is consistent with the definition in \eref{eq:conjugate_force} and the LF transform \eqref{eq:convex_conj}.}
In the dual (force-based) formulation, the flow rule becomes
\begin{equation} \label{eq:GSM_dual_normality_revised}
\dot{\internalvariables} = \partialb_{\conjugateforce} \dissipationpotential(\conjugateforce; \dot{\lagrangestrain},\abstemperature ) \ ,
\end{equation}
as in the DP formulation; \ADD{ however, in a GSM formulation, the pseudo-potential $\dissipationpotential$ can only depend on $\{ \conjugateforce; \dot{\Eb},\abstemperature\}$ rather than on the full auxiliary state $\Zc$ due to \eref{eq:dual_dissipation_potential}, and must be interpreted as the dual potential in the sense of convex conjugacy \eqref{eq:convex_conj}.  }

With the dual formulation. the dissipation requirement \eref{eq:dissipation_gsm} becomes
\begin{equation} \label{eq:dual_dissipation_gsm}
\dissipation
= \partialb_{\dot{\lagrangestrain}} \dissipationpotential^* \cdot \dot{\lagrangestrain}
\ADD{+} \partialb_{\dot{\internalvariables}} \dissipationpotential^* \cdot \dot{\internalvariables}
= -\partialb_{\dot{\lagrangestrain}} \dissipationpotential \cdot \dot{\lagrangestrain}
\ADD{+} \conjugateforce \cdot \partialb_\conjugateforce \dissipationpotential  \ge 0
\end{equation}
by virtue of the LF transform \eref{eq:gsm_LF}.

\ADD{
\begin{remark}
The dissipation requirement \eref{eq:dual_dissipation_gsm} presents an implementation challenge even in a simplified sufficient condition were \eref{eq:dual_dissipation_gsm} is taken to be two independent inequalities, since it is difficult to ensure one derivative , $\partialb_{\dot{\lagrangestrain}} \dissipationpotential \cdot \dot{\lagrangestrain}$,  being strictly non-positive and the other, $\conjugateforce \cdot \partialb_\conjugateforce \dissipationpotential$, non-negative.
In fact this implies saddle at the reference state.%
\footnote{With $\xb = \dot{\Eb}$ and $\yb = \dot{\internalvariables}$: $\partialb_{\xb} \dissipationpotential(\xb,\yb) \cdot \xb \ge \dissipationpotential(\xb,\yb) - \dissipationpotential(\zerob,\yb)$ and $\dissipationpotential$ is minimized at $\xb=\zerob$; and conversely $\partialb_{\yb} \dissipationpotential(\xb,\yb) \cdot \yb \ge \dissipationpotential(\xb,\yb) - \dissipationpotential(\xb,\zerob)$ and $\dissipationpotential$ maximized at $\xb=\zerob$.}
Ideally we need a general function representation for convex-concave function of two tensor arguments, which currently we are not aware of;
however, a number of options are apparent:%
\footnote{The use of neural network representations of Euler homogeneous functions  \cite{polyakov2024generalized}, with positive or negative degree,  may also have some utility in satisfying the physical requirements, where if $\dissipationpotential(k \xb) = k^n \dissipationpotential(\xb)$ then $\partialb_{\xb} \dissipationpotential(\xb) = n \dissipationpotential(\xb)$), together with the use of the representataion $ \dissipationpotential = \| \xb \|^n \NN(\xb/\|\xb\|)$ where $\NN$ is on the hypersphere.}
\begin{enumerate} \renewcommand{\labelenumi}{(\alph{enumi})}
\item enforce the Biot relation \eqref{eq:GSM_Biot_revised}, e.g. via a penalty as in \cref{rosenkranz2023comparative},  but this is a soft constraint and introduces a hyperparameter.
\item omit the dependence of $\dissipationpotential$ on the rate $\dot{\lagrangestrain}$, as in \cref{flaschel2025convex} i.e. $\dissipationpotential = \dissipationpotential(\conjugateforce)$, but this may limit the expressiveness of the framework.
\item make $\dissipationpotential$ additively separable $\dissipationpotential = \dissipationpotential(\conjugateforce) + \varphi(\Ic_{\dot{\Eb}})$, an extension of (b), or multiplicatively $\dissipationpotential = \dissipationpotential(\conjugateforce)  \varphi(\Ic_{\dot{\Eb}})$, analogous to \cref{fuhg2024polyconvex}.
\end{enumerate}
\end{remark}
}

\ADD{
Since the canonical function $f(x,y)$ for axis-aligned convexity-concavity is the additively separable difference of two convex functions $f = x^2 - y^2$. i.e convex for x and concave for y where $x^2$ and $y^2$ are convex,  we elect to represent $\dissipationpotential_{\text{GSM}}$ as:
\begin{equation}
\dissipationpotential_\text{GSM} = \varphi_{\conjugateforce}(\conjugateforce) - \varphi_{\dot{\Eb}}(\dot{\Eb})
\end{equation}
where both functions $\varphi_{\conjugateforce}$ and $\varphi_{\dot{\Eb}}$ are convex in their arguments.
This can be seen as an extension of choice (b).
}

\paragraph{Metriplectic (MP)}
The \emph{metriplectic} (MP) \cite{morrison1984bracket,morrison1986paradigm}  and the related \emph{General} \emph{Equation} \emph{for} \emph{Non-Equilibrium}  \emph{Reversible-Irreversible} \emph{Coupling} (GENERIC) \cite{ottinger1997dynamics,grmela1997dynamics}   formalisms are not specifically tailored to constitutive modeling.
Instead, these frameworks are meant to be a general representation of non-equilibrium thermodynamics \cite{grmela1997dynamics} across a variety of scales; nevertheless, a few authors \cite{mielke2011formulation,schiebl2021structure,hutter2011formulation,hutter2008continuum,hutter2008finite,hutter2015statistical} have applied them to constitutive modeling of inelastic materials.

\begin{remark}
In addition to a general set of states $\zb$, the metriplectic formalism has four structures:
a (symplectic) Hamiltonian operator $\Lb(\zb)$,
a (metric) dissipation operator $\Mb(\zb)$,
an energy potential $e$, and an entropy potential $s$.
The dynamical system is simply
\begin{equation}
\dot{\zb}
= \Lb \partialb_\zb e + \Mb \partialb_\zb s \ ,
\end{equation}
or more generally \cite{grmela2025rheological}
\begin{equation}
\dot{\zb}
= \Lb \partialb_\zb e + \left. \partialb_{\zb^*} \Xi(\zb,\zb^*) \right|_{\zb^* = \partialb_\zb s} \ ,
\end{equation}
where $\Xi$ is a generalized dissipation potential.
The two components drive different aspects of the dynamics.
The energy potential $e$ generates Hamiltonian dynamics, such that:
\begin{equation} \label{eq:e_flow}
\dot{e} = \partialb_{\zb} e \cdot \dot{\zb}
= \partialb_{\zb} e \cdot  \Lb \, \partialb_{\zb} e \equiv 0 \ ,
\end{equation}
since $\Lb = -\Lb^T$ is skew-symmetric,
while the dissipation potential $s$ generates an entropy-producing flow:
\begin{equation} \label{eq:s_flow}
\dot{s} = \partialb_{\zb} s \cdot \dot{\zb}
=  \partialb_{\zb} s \cdot  \Mb \, \partialb_{\zb} s
\ge 0 \ ,
\end{equation}
given that \ADD{ $\Mb = \Mb^T$ } is symmetric, positive definite.
The degeneracy conditions
\begin{equation} \label{eq:mp_degeneracy}
\Lb \partialb_\zb s = \zerob
\quad \text{and} \quad
\Mb \partialb_\zb e = \zerob \ ,
\end{equation}
are necessary for \eref{eq:e_flow} and \eref{eq:s_flow} to hold.
Note the non-decreasing $s$ can play the role of a Lyapunov function on the dynamics \cite{grmela2025rheological} and this framework is typically described in terms of a Poisson bracket (with an attendant Jacobi identity) \cite{grmela2018generic}, which we omit for consistency with other formulations.
\end{remark}

As in the DP and GSM formulations, this MP formulation relies on two potentials doing separate tasks; here, it is Hamiltonian dynamics with dissipation added to describe relaxation to a thermodynamic equilibrium.
Following Ziegler's discussion \cite{ziegler1972thermomechanics}, suggesting a Hamiltonian micro-system evolving with macro state variables, we apply the MP framework to the dynamics of the internal states:
\begin{equation} \label{eq:MP_flow}
\dot{\internalvariables} =  \Lb \underbrace{\partialb_{\internalvariables} \internalenergy}_{\gb} + \Mb \underbrace{\partialb_{\internalvariables} \dissipationpotential}_{\hb}
\end{equation}
Based on fact that $s$ is associated with the \emph{Massieu} potential/\emph{free entropy}  \cite{callen1993thermodynamics,hernandez2023port,mielke2011formulation}, we identify $s$ with the entropy $\entropy$ as modeled by $\dissipationpotential$ and $e$ with the internal energy $\internalenergy$ as the stress potential, which is consistent with the definition of $\freeenergy$ \eqref{eq:helmholtz} with fixed temperature.
With the definition of the free energy \eqref{eq:helmholtz} and
\begin{equation}
\conjugateforce \equiv - \partialb_\internalvariables \freeenergy
= - \partialb_\internalvariables \internalenergy + \abstemperature \partialb_\internalvariables \entropy
= -\gb + \abstemperature \hb
\end{equation}
the dissipation requirement  \eqref{eq:dissipation_req} for MP reduces to:
\begin{equation} \label{eq:mp_dissipation}
\dissipation
= \conjugateforce \cdot \dot{\internalvariables}
=  \ADD{(\abstemperature \hb - \gb)}  \cdot
\ADD{
( \Lb \gb + \Mb \hb )
}
= \ADD{\abstemperature \, \hb\cdot\mathbf{M}\hb } \ge  0
\end{equation}
after using the MP flow \eqref{eq:MP_flow}, \ADD{the properties of $\Lb$ and $\Mb$,} and the degeneracy conditions \eqref{eq:mp_degeneracy}.
Thermodynamic admissibility is therefore ensured independently of the functional forms of $\freeenergy$ and $\dissipationpotential$ other than the degeneracy conditions \eqref{eq:mp_degeneracy} and the operator symmetries.%
\footnote{The form of the MP dynamics is reminiscent of flow due to quasi-quadratic potentials \cite{onsager1931reciprocalI,onsager1931reciprocalII,ottosen2005mechanics}, see \aref{appendix:ISM} for more details.}

\paragraph{Implicit standard materials (ISM)}
As mentioned, the ISV framework leaves the evolution map $\flow$ largely unrestricted beyond the dissipation requirement \eqref{eq:dissipation_req}.
A structured intermediate generalization is provided by implicit standard materials (ISM) \cite{de1992generalisation,bodoville2001implicit}, in which the evolution is not given by an explicit map, such as $\dot{\internalvariables}=\flow(\lagrangestrain,\internalvariables)$, but instead by an implicit force--flux relation generated by a \emph{bipotential} $\beta$:
\begin{equation}
\beta(\conjugateforce,\dot{\internalvariables})  \;\ge\;  \conjugateforce \cdot \dot{\internalvariables}.
\end{equation}
Admissible pairs $(\conjugateforce,\dot{\internalvariables})$ lie on the monotone graph defined by equality in this generalized Fenchel inequality.
This construction guarantees non-negative dissipation by design, while remaining more general than the DP, GSM, and MP structures.
In particular, DP and GSM correspond to distinct convex choices of the bipotential, while MP can be interpreted as a quadratic metric specialization (see \aref{appendix:ISM}).
\ADD{The flow rules $\conjugateforce \in \partialb_{\dot{\internalvariables}} \dissipationpotential^*$ and $\dot{\internalvariables} \in \partialb_{\conjugateforce} \dissipationpotential$ extend the GSM flow rules, \eref{eq:GSM_Biot_revised} and \eref{eq:GSM_dual_normality_revised}, by allowing any values within the subdifferentials of the dissipation potentials \cite{berga1994elastoplastic,de2002implicit}}
\ADD{In particular, GSM is recovered if $\beta = \dissipationpotential + \dissipationpotential^*$ (refer to the Remark at the beginning of this section).}

To the authors' knowledge, a physics-augmented neural network (PANN) implementation of the full ISM framework has not yet been reported.
Although ISM provides a unifying and highly general convex-analytic structure, embedding a general bipotential \ADD{and subdifferentials} directly into neural potential-based architectures introduces additional modeling and optimization challenges.
For this reason, the present work focuses on the DP, GSM, and MP subclasses, leaving a PANN realization of ISM for future work.

\begin{table}[h]
\centering
\begin{tabular}{|l|c|ccc|c|}
\hline
Feature & ISV & DP & GSM & MP & \ADD{ISM}  \\
\hline
\hline
local equilibrium                & assumed & assumed & not assumed & not assumed  & not assumed \\
\hline
free energy                      &\textbf{required}  &\textbf{required}&\textbf{required} & allowed & \textbf{required} \\
\quad convexity                  &allowed            &allowed  &\textbf{required}   &  allowed  & \textbf{required} \\
\quad internal variables         &included           &included &included & allowed     & included\\
\hline
dissipation potential            &allowed            &\textbf{required} &\textbf{required}  & \textbf{required} & \textbf{required}  \\
\quad convexity                  &allowed            &allowed  &\textbf{required}   & metric  &\textbf{required}\\
\quad dual potential             &no                 &no       &\textbf{yes} & no     & \textbf{yes}\\
\quad dependence on state        &free               &free     &\textbf{restricted} & free    & \textbf{restricted}  \\
\quad normality/associated flow  &allowed            &allowed  &\textbf{required}   & allowed & allowed \\
\hline
\end{tabular}
\caption{Summary of distinctions between the constitutive frameworks.}
\label{tab:formulation_distinctions}
\end{table}

\section{Implementation} \label{sec:implementation}

\begin{figure}
\centering
\includegraphics[width=0.65\linewidth]{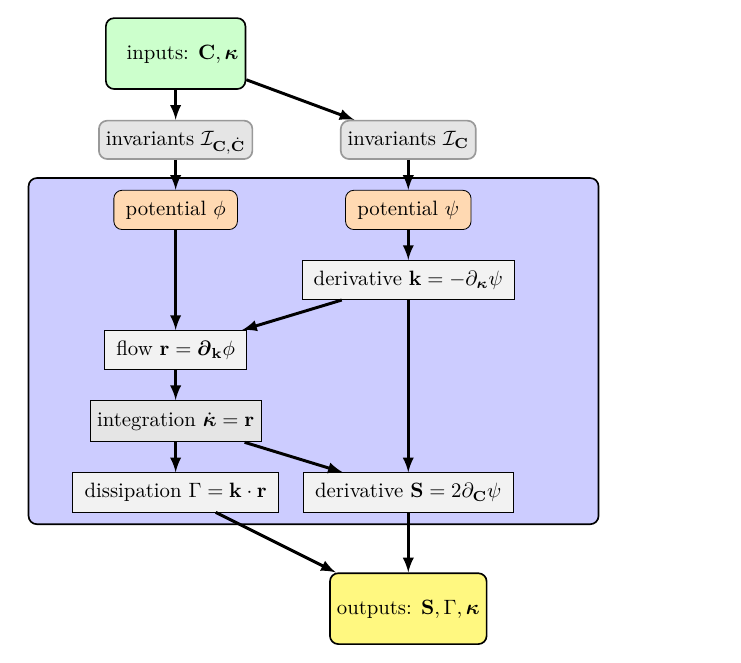}
\caption{Schematic of the DP framework}
\label{fig:schematic}
\end{figure}

We implement our interpretations of potential-based DP, GSM, and MP within a common framework consisting of: an energy potential and a dissipation potential, with corresponding stress and flow rules.
\fref{fig:schematic} illustrates the common framework.
We use the flexible \emph{input specific neural network} (ISNN) \cite{jadoon2026input} to implement the potentials and a neural ODE \cite{chen2018neural,dupont2019augmented} approach for the evolution.
The implementation differences are summarized in \tref{tab:implementations} at the end of this section.
In the following, we drop the temperature dependence for convenience.

\paragraph{Invariance}
\ADD{All potentials, being scalar-valued functions, need to be functions of scalar invariants.}
Hence, we assume all the ISVs are \ADD{independent} scalar invariants, although other approaches could be taken \cite{jones2025physics}.
For the observable inputs, we assume isotropy and use the Cayley-Hamilton principal invariants for deformation measure $\Cb$ and  Landau power invariants%
\footnote{\ADD{We have used the Landau invariants $\tr \dot{\Cb}^k, k \in \{1,2,3\}$ in the past \cite{jones2022neural} but $\tr \dot{\Cb}^3$ is not convex, which is required by our GSM implementation; hence we substitute $\tr\dot{\Cb}^4$ \cite{rosenkranz2024viscoelasticty} and assume this set has the same expressive power due the Cayley-Hamilton theorem through the identity:
$\operatorname{tr}(\Cb^4) = \frac{1}{6}\operatorname{tr}(\Cb)^4 - \operatorname{tr}(\Cb)^2\operatorname{tr}(\Cb^2) + \frac{1}{2}\operatorname{tr}(\Cb^2)^2 + \frac{4}{3}\operatorname{tr}(\Cb)\operatorname{tr}(\Cb^3)$.}}
for the rate $\dot{\Cb}$:
\begin{equation} \label{eq:invariants}
\begin{array}{llll}
\multicolumn{2}{c}{\overbrace{\hspace{1in}}^{\Ic_{\Cb}}} &
\multicolumn{2}{c}{\overbrace{\hspace{0.7in}}^{\Ic_{\dot{\Cb}}}} \\
I_1 &= \tr \Cb            &\quad I_4 &= \tr \dot{\Cb} \\
I_2 &= \tr \Cb^*          &\quad I_5 &= \tr \dot{\Cb}^2  \\
I_3 &= \sqrt{\det \Cb}    &\quad I_6 &= \ADD{\tr \dot{\Cb}^4 }\\
\multicolumn{4}{c}{\underbrace{\hspace{2in}}_{\Ic_{\Cb,\dot{\Cb}}}}  \\
\end{array}
\end{equation}
where $\Cb^* = \det(\Cb) \Cb^{-T}$ and $I_3$ is $\det\defgrad$, not the conventional $\det\Cb$, for polyconvexity considerations.
We remark that other invariants are possible \cite{fuhg2024stress}%
\footnote{The choice of invariants is, in some sense, a feature engineering exercise, as \cref{fuhg2024stress} explores.
The usual Cayley-Hamilton or Landau invariants can be shifted, scaled, or otherwise transformed, such as taking roots so that they are dimensionally homogeneous
\cite{holthusen2025generalized,holthusen2026complement};
however, the invariants must be permutationally symmetric in the stretches; convexity is preserved by positive-definite affine transformations; and different functional forms can lead to different thermodynamic consequences.}
and that this can be extended to anisotropy via the choice of appropriate invariants \cite{fuhg2022learning,patel2025general}.
Since the two arguments, $\Cb$ and $\dot{\Cb}$, are related by a time derivative, these 6 invariants are sufficient (see appendix of  \cref{jones2022neural}.
To maintain invertibility \ADD{and therefore the ability to compute the adjugate and its trace in the reference configuration},  we use the stretch $\Cb$ instead of the strain $\lagrangestrain$.
The  basis elements $\Bc$ result from derivatives of the invariants: $\Bc_{\Cb} = \{ \partialb_{\Cb} I_a \}_{a=1}^{6}$ and  $\Bc_{\dot{\Cb}} = \{ \partialb_{\dot{\Cb}} I_a\}_{a=1}^{6}$.

Note, with this choice of invariants, the GSM dissipation \eqref{eq:dual_dissipation_gsm} can be written as:
\begin{eqnarray}
\dissipation
&=& -\partialb_{\dot{\lagrangestrain}} \dissipationpotential \cdot \dot{\lagrangestrain}
+\conjugateforce \cdot \partialb_\conjugateforce \dissipationpotential  \\
&=& - \ADD{\tfrac{1}{2}}\partial_{\Ic_{\dot{\Cb}}} \dissipationpotential \; \partialb_{\dot{\Cb}} \Ic_{\dot{\Cb}} \cdot \dot{\Cb}
+\conjugateforce \cdot \partialb_\conjugateforce \dissipationpotential
\ge 0    \nonumber
\end{eqnarray}
and further simplified  with the aid of the identities:
\begin{equation}
\dot{\Cb} \cdot \partialb_{\dot{\Cb}} \tr \dot{\Cb}   =   \tr \dot{\Cb}   =   I_4 \ , \quad
\dot{\Cb} \cdot \partialb_{\dot{\Cb}} \tr \dot{\Cb}^2 =   2 \tr \dot{\Cb}^2 = 2 I_5 \ADD{\ge 0} \ , \quad
\ADD{\dot{\Cb} \cdot \partialb_{\dot{\Cb}} \tr \dot{\Cb}^4 =   4 \tr \dot{\Cb}^4 = 4 I_6 \ge 0} \ .
\end{equation}

\paragraph{Potentials}
Based on the theory in \sref{sec:theory}, the energy and dissipation potentials have different input arguments depending on the particular framework.
The dissipation potential (DP) framework has an (equilibrium) free energy
\begin{equation}
\freeenergy =  \hat{\freeenergy}(\Ic_\Cb, \internalvariables) \ ,
\end{equation}
We ensure that $\freeenergy$ is polyconvex \cite{ball1976convexity} by having $\freeenergy$ be convex and monotone in $I_1$, \ADD{$I_2$}, and convex in $I_3 >0$ \cite{schroder2003invariant,mielke2005necessary,schroder2010anisotropie}.%
\footnote{Polyconvexity with respect to the principal invariants of the deformation gradient is ensured by these conditions on $\freeenergy(I_i)$ since $I_3=J$ is already one of the principle invariants \eqref{eq:invariants}.}
We allow the DP dissipation potential to depend on the deformation and rate invariants as well as the internal state and conjugate force%
\footnote{Convexity of $\dissipationpotential$ requires convexity of the invariants with respect to the eigenvalues of the input tensors and monotonicity of the potential with respect to the these invariants.}
\begin{equation}
\dissipationpotential_{\text{DP}} = \dissipationpotential_{\text{DP}}(\conjugateforce,\Ic_{\Cb,\dot{\Cb}},\internalvariables) \, .
\end{equation}
The GSM framework prescribes the same free energy dependence as DP and a dissipation potential
\begin{equation}
\dissipationpotential_{\text{GSM}} = \dissipationpotential_{\text{GSM}}(\conjugateforce,\Ic_{\dot{\Cb}}) \ ,
\end{equation}
only dependent on the conjugate force and the rate invariants.%
\footnote{We construct $ \phi_{\mathrm{GSM}} = \phi_k(\kb)-\phi_{\dot{\Cb}}(I_{\dot{\Cb}}), $ with $\phi_k$ and $\phi_{\dot{\Cb}}$ convex and minimized at the origin.
Hence, $ \kb\cdot\partial_{\kb}\phi_k \ge 0, \qquad \dot{\Cb}:\partial_{\dot{\Cb}}\phi_{\dot{\Cb}} \ge 0, $
and therefore $\Gamma = -\partial_{\dot{\Cb}}\phi_{\mathrm{GSM}}:\dot{\Cb} + \kb\cdot\partial_{\kb}\phi_{\mathrm{GSM}} = \dot{\Cb}:\partial_{\dot{\Cb}}\phi_{\dot{\Cb}} + \kb\cdot\partial_{\kb}\phi_k \ge 0$, as opposed to insuring $\sum_a I_a\,\partial_{I_a}\phi \ge 0$.
}
Our version of the MP framework uses an internal energy potential:
\begin{equation} \label{eq:mp_potentials}
\internalenergy = \internalenergy(\Ic_{\Cb},\internalvariables) \ ,
\end{equation}
with the same inputs as free energy in the other frameworks, while the dissipation potential:
\begin{equation}
\dissipationpotential_{\text{MP}} = \dissipationpotential_{\text{MP}}(\internalvariables,\Ic_{\Cb,\dot{\Cb}}) \ ,
\end{equation}
only depends on the internal state and the full set of invariants, but not the conjugate force.

\paragraph{Stress}
The DP implementation uses the standard (local equilibrium) stress rule:
\begin{equation}
\pkstress = 2 \partialb_{\Cb} \freeenergy \, .
\end{equation}
The MP implementation uses the equivalent stress rule:
\begin{equation}
\pkstress = 2 \partialb_{\Cb} \internalenergy \ ,
\end{equation}
while our implementation of the GSM model uses an Onsager-like rule:
\begin{equation}
\pkstress = 2 \partialb_{\Cb} \freeenergy - 2 \partialb_{{\dot{\Cb}}} \dissipationpotential_{\mathrm{GSM}}
\end{equation}

We impose that the stress is zero in the reference state:
\begin{equation} \label{eq:ref_state}
\pkstress = \zerob \quad \text{if} \quad (\Cb = \Ib, \dot{\Cb} = \zerob,  \internalvariables = \zerob)
\end{equation}
using normalization schemes.%
\footnote{ \ADD{Note the imposition of a stress-free reference is common but not a necessary condition.} }
For the DP and MP formulations, we shift the free energy
\begin{equation} \label{eq:dp_psi}
\freeenergy= \hat{\freeenergy}(\Ic_{\Cb},\internalvariables)  -p_\freeenergy\,(I_3-1) \ .
\end{equation}
by\footnote{
This normalization results from:
$
\pkstress
= 2 \left( (\partial_{I_1} \freeenergy + I_1 \partial_{I_2} \freeenergy) \, \Ib
- \partial_{I_2} \freeenergy \, \Cb
+  I_3 \partial_{I_3} \freeenergy \, \Cb^{-1}  \right) \ ,
$
restricted to the reference configuration $\Cb=\Ib$.
}
\begin{equation}\label{eq:o_def_internal}
p_\freeenergy(\internalvariables) =
\left. \left[
2\partial_{I_1}\hat\freeenergy + 4\partial_{I_2}\hat\freeenergy + \partial_{I_3}\hat\freeenergy
\right] \right|_{\Ic_{\Cb}=\zerob} \ .
\end{equation}
For the GSM formulation, this shift of the free energy is needed in addition to a similar correction of the raw dissipation potential:
\begin{equation}
\dissipationpotential_{\text{GSM}} = \hat{\dissipationpotential}_{\text{GSM}}(\conjugateforce,\Ic_{\dot{\Cb}}) - p_\dissipationpotential(\conjugateforce)\, I_4 \ ,
\end{equation}
where the normalization component:
\begin{equation}\label{eq:b_def}
p_\dissipationpotential(\conjugateforce) \, I_4
=
\left.\partial_{I_4}\hat\dissipationpotential_{\mathrm{GSM}}(\conjugateforce,\Ic_{\dot{\Cb}})\right|_{\conjugateforce=\mathbf 0 ,\Ic_{\dot{\Cb}}=\mathbf 0} \, I_4 \ ,
\end{equation}
is affine in $I_{4}$.%
\footnote{These corrections are of the general form $\phi(\xb) = \hat{\phi}(\xb) - \hat{\phi}(\xb_0) - \left. \partialb_{\xb} \hat{\phi} \right|_{\xb_0} \cdot (\xb-\xb_0)$.}
This correction is expressed in terms of $I_4=\tr\dot{\Cb}$ since $\Bb_4 = \partial_{\dot{\Cb}} I_4=\Ib$ remains nonzero at $\dot{\Cb}=\zerob$, whereas the other basis elements (derivatives of higher-order rate invariants) vanish at the quiescent reference state, making $I_4$ the only of the chosen invariants capable of removing a linear drift contribution in $\partial_{\dot{\Cb}}\dissipationpotential_{\mathrm{GSM}}$ and thereby enforcing zero stress in the reference state.
\footnote{
Note, $I_4 = \tr \dot{\Cb} = 0$ does not imply $\dot{\Cb}=\zerob$, as opposed to  $I_5 = \| \dot{\Cb} \|^2$ which \ADD{could} also be used in the normalization.
}

\paragraph{Flow}
In both the DP and GSM implementations, the flow $\dot{\internalvariables} = \flow$ is given by the gradient of the dissipation potential with respect to the conjugate force:
\begin{equation}
\flow =  \partialb_{\conjugateforce} \dissipationpotential \ ,
\end{equation}
while the MP implementation uses metriplectic dynamics for the evolution of the internal states:
\begin{equation}
\flow = \Lb \partialb_{\internalvariables} \internalenergy +  \Mb \partialb_{\internalvariables} \dissipationpotential \ .
\end{equation}
Dissipation is associated with flow, which implies:
\begin{equation}
\flow = \zerob \quad \text{if} \quad \conjugateforce = \zerob \, ,
\end{equation}
\ADD{which like the zero stress condition \eqref{eq:ref_state} comes from the expectation that fluxes should be zero at a  quiescent reference state and ensures the appropriate potential is minimized with respect to respective inputs.}
To enforce this, we employ shifts analogous to those used in the stress normalization schemes:
\begin{equation}
\ADD{\dissipationpotential} = \hat{\dissipationpotential} - \left. \partialb_{\conjugateforce} \hat{\dissipationpotential} \right|_{\conjugateforce=\zerob} \cdot \conjugateforce
\end{equation}

To maintain parity (in terms of NN parameters) between the MP implementation and the others, we employ fixed $\Lb$ and $\Mb$:
\begin{equation}
[ \Lb ]_{ij} = \begin{cases}
1   & i>j \\
-1  & i<j \\
0   & \text{else} \\
\end{cases}
\quad \text{and} \quad
[ \Mb ]_{ij} = \begin{cases}
2   & i=j \\
1   & \text{else} \\
\end{cases}
\end{equation}
that do not bias the evolution toward particular components of the internal state vector.
Others have used fixed operators \cite{hernandez2021structure}, albeit adapted to the particular physics, \ADD{and full neural network representation of the MP formulation  is generally known to be redundant \cite{gruber2025efficiently}}.
Learnable operators \cite{gruber2025efficiently} may offer representational advantages and are easily implemented, but they are out-of-scope for this work.

The MP degeneracy conditions \eqref{eq:mp_degeneracy} can be embedded by transforming the operators with projections \cite{gruber2025efficiently}
\begin{equation}
\Lb'(\internalvariables) = \Pb_\hb \Lb \Pb_\hb
\quad \text{and} \quad
\Mb'(\internalvariables) = \Pb_\gb \Mb \Pb_\gb
\end{equation}
where
\ADD{$\gb\equiv \partial_{\internalvariables}\internalenergy$
and
$\mathbf{h}\equiv \partial_{\boldsymbol{\kappa}}\dissipationpotential$,} and
\begin{equation}
\Pb_{\gb} = \Ib - \frac{\gb}{\| \gb \|} \otimes \frac{\gb}{\| \gb \|}
\end{equation}
and likewise for $\Pb_\hb$.

\begin{remark}
It may be better to parameterize the operators such as $\Pb_\gb$ as  $\Pb'_{\gb} = (\gb\cdot\gb) \Ib - \gb \otimes \gb$ to handle cases were $\gb \to \zerob$ but $\Pb'_{\gb}$ is not a projection.
In fact, Birtea \cite{birtea2009asymptoic}, as an alternative to the method we adopt \cite{gruber2025efficiently}, uses $\Pb'$ as the operator $\Mb'$ thus subverting the need to learn or choose $\Mb$.
\end{remark}

\paragraph{Dissipation}
To satisfy dissipation, the DP formulation employs a dissipation potential
$\dissipationpotential(\conjugateforce)$ that is convex in the conjugate force,
non-negative, and minimized at $\conjugateforce = \boldsymbol{0}$. Under these conditions,
\begin{equation}
\dissipation
= \conjugateforce \cdot \flow
= \conjugateforce \cdot \partialb_{\conjugateforce} \dissipationpotential
\ge 0 \ .
\end{equation}
For the GS formulation, the dual dissipation potential  $\dissipationpotential^*(\dot{\internalvariables},\dot{\lagrangestrain})$  must be convex in its kinematic rate arguments and minimized at  $(\dot{\lagrangestrain},\dot{\internalvariables}) = (\boldsymbol{0},\boldsymbol{0})$,  in part due to the dissipation requirement \eref{eq:dual_dissipation_gsm}.
Akin to the polyconvexity requirements on the free energy, $\dissipationpotential_\text{GSM}(\conjugateforce, \dot{\Cb})$ is monotonic and concave in the rate invariants since it needs to be concave in the rate of deformation, while it is also convex in conjugate force since $\conjugateforce$ is simply a collection of scalar invariants.
The MP formulation satisfies the dissipation requirement by virtue of the symmetries of the operators $\Lb$ and $\Mb$ together with the degeneracy conditions \eqref{eq:mp_degeneracy}.

\begin{table}[h!] \renewcommand{\arraystretch}{1.5}
\centering
\begin{tabular}{|l|c|c|c|}
\hline
component & DP & GSM & MP \\
\hline
\hline
energy                  & $\freeenergy    (\Ic_{\Cb}, \internalvariables)$
& $\freeenergy    (\Ic_{\Cb}, \internalvariables)$
& $\internalenergy(\Ic_{\Cb}, \internalvariables)$  \\
\quad minimum at        & $\Ic_{\Cb} = [3,3,1]$
& $\Ic_{\Cb} = [3,3,1]$
& $\Ic_{\Cb} = [3,3,1]$ \\
\quad monotone \& convex            & $I_1,I_2$
& $I_1,I_2$
& $I_1,I_2$ \\
\quad convex & $I_3$
& $I_3$
& $I_3$  \\
\quad free              & $\internalvariables$
& $\internalvariables$
& $\internalvariables$ \\
\hline
dissipation             & $\dissipationpotential(\conjugateforce, \Ic_{\Cb,\dot{\Cb}}, \internalvariables)$
& $\dissipationpotential(\conjugateforce, \Ic_{\dot{\Cb}})$
& $\dissipationpotential(\Ic_{\Cb,\dot{\Cb}}, \internalvariables)$  \\
\quad minimum at     & $\conjugateforce = \zerob$
& $\conjugateforce = \zerob$, $\Ic_{\dot{\Cb}} = 0$
& -- \\
\quad monotone \& convex  &  --
&  $\Ic_{\dot{\Cb}}$
& --  \\
\quad convex            & $\conjugateforce$
& $\conjugateforce$
& -- \\
\quad free              & $\Ic_{\Cb,\dot{\Cb}}$, $\internalvariables$
&  --
& $\Ic_{\Cb,\dot{\Cb}}$, $\internalvariables$ \\
\hline
stress                  & $\pkstress = 2 \partialb_{\Cb} \freeenergy$
& $\pkstress = 2 \partialb_{\Cb} \freeenergy - 2 \partialb_{{\dot{\Cb}}} \dissipationpotential$
& $\pkstress = 2 \partialb_{\Cb} \freeenergy$ \\
conjugate force         & $\conjugateforce = -\partialb_{\internalvariables} \freeenergy$
& $\conjugateforce = -\partialb_{\internalvariables} \freeenergy$
& $\conjugateforce = -\partialb_{\internalvariables} \internalenergy + \abstemperature  \partialb_{\internalvariables} \dissipationpotential$ \\
flow                    & $\flow = \partialb_{\conjugateforce} \dissipationpotential$
& $\flow = \partialb_{\conjugateforce} \dissipationpotential$
& $\flow = \Lb \partialb_{\internalvariables} \internalenergy +  \Mb \partialb_{\internalvariables} \dissipationpotential$ \\
dissipation             & $\conjugateforce \cdot \flow$
& $\conjugateforce \cdot \flow
-\partial_{\Ic_{\dot{\Cb}}} \dissipationpotential \, \partialb_{\dot{\Cb}} \Ic_{\dot{\Cb}} \cdot \dot{\Cb}$
& $\conjugateforce \cdot \flow$ \\
\hline
\end{tabular}
\caption{Implementation comparison. Note: \ADD{(a) the free energy $\freeenergy$ is polyconvex, (b) the dissipation potential for each formulation is sufficiently convex to ensure dissipation, i.e. convex with respect to the gradient input, e.g. $\conjugateforce$ for DP, and (c)} $\abstemperature=1$ is assumed for MP.}
\label{tab:implementations}
\end{table}

\section{Data} \label{sec:data}

To generate training and testing data, we use three different inelastic representative volume elements (RVEs): an alloy, a composite, and a polycrystal.
We use a single realization of each and run 256 trajectories of cyclic uniaxial loading  using:
\begin{equation} \label{eq:F_samples}
\defgrad(t) = \Ib + \sum_i a_i \sin(\omega_i t) \, \eb_i \otimes \eb_i \ ,
\end{equation}
where the amplitudes $a_i$ and frequencies $\omega_i$ are drawn from uniform distributions.
This deformation is applied via periodic boundary conditions.
Each trajectory has 200 steps, and we observe only the 11\ADD{-component} of the resulting stress trajectories in emulation of typical uniaxial experiments \ADD{under the assumption of homogeneous deformation with actuator force and displacement measurements}.%
\footnote{The invariant formulation we adopt generally informs the other components.}
\ADD{The data was split, by trajectory into 80/10/10 training, in-training testing, and held out validation sets.}
\ADD{We chose RVE data-generating models so that there would not be a simple closed form to the response that would potentially distort the formulation comparisons.}

\subsection{AlSi$_{10}$Mg elastic-plastic alloy (EP)}
\begin{figure}
\centering
\begin{subfigure}[c]{0.35\linewidth}
\includegraphics[width=0.95\linewidth]{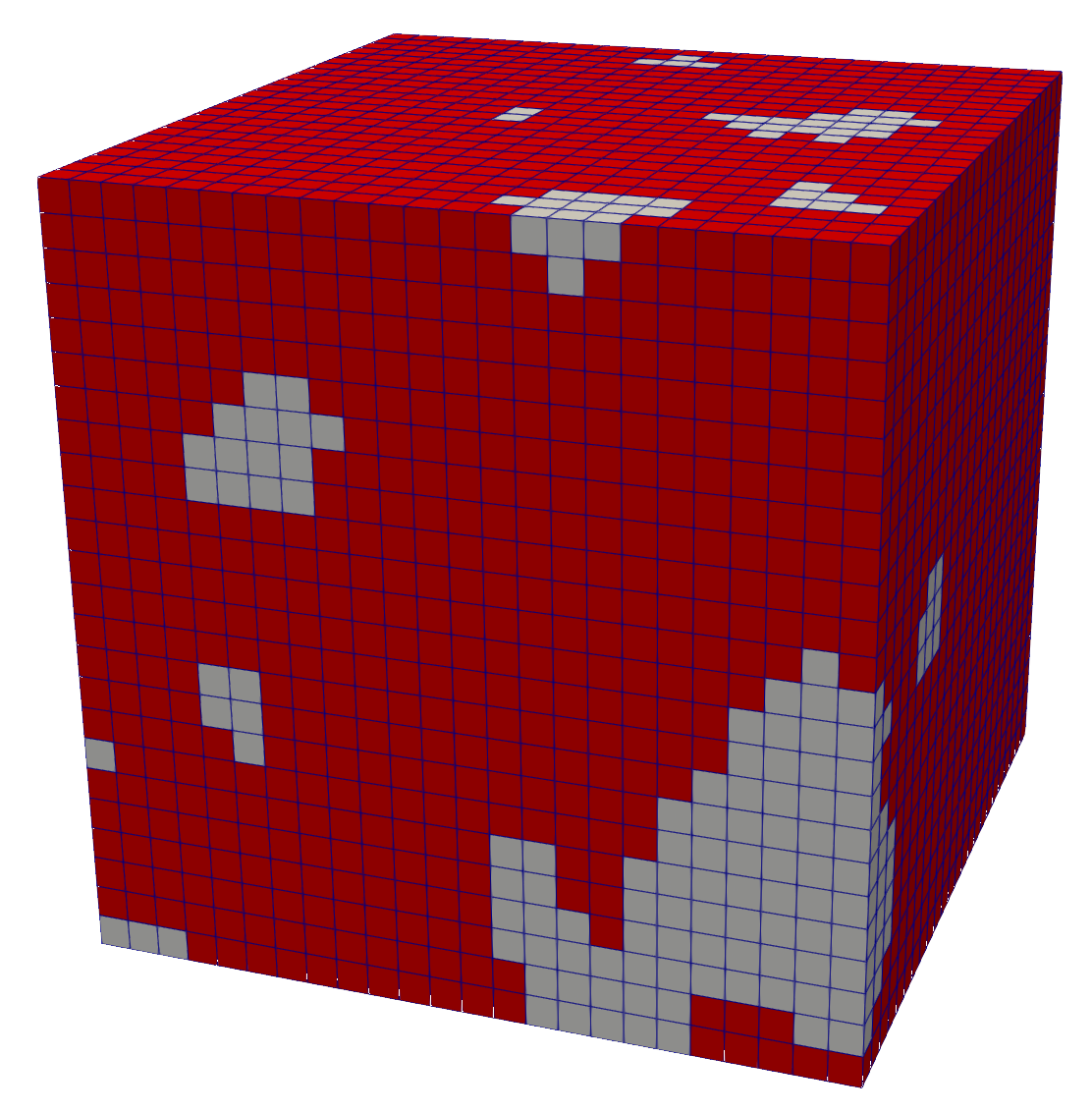}
\end{subfigure}
\begin{subfigure}[c]{0.5\linewidth}
\includegraphics[width=0.95\linewidth]{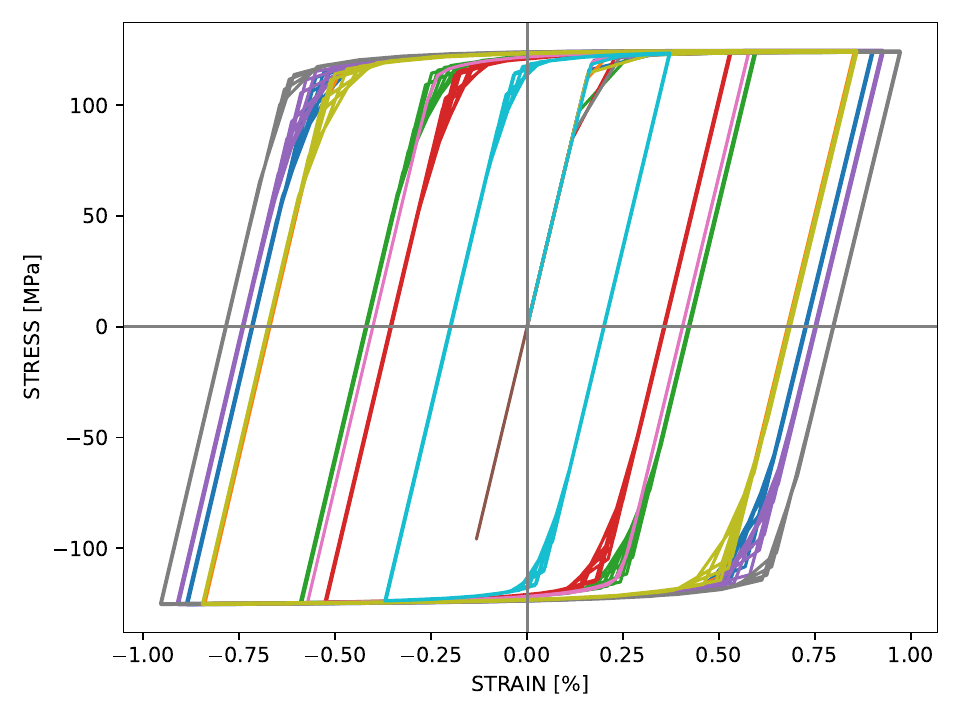}
\end{subfigure}
\caption{AlSi$_{10}$Mg alloy. \ADD{ (a) representative  volume element (red: Al, gray:Si), (b) representative hysteresis curves of a random ensemble where realizations are distinguished by color. }}
\label{fig:alloy}
\end{figure}

AlSi$_{10}$Mg is a common additive manufacturing material.
We idealize it as an alloy of elastic Si and elastic-plastic Al, where the alloy structure was generated using a Gaussian random field \cite{liu2019advances} on a $25^3$ structured mesh.
Both phases use a St. Venant elastic response, and the Al phase employs a J2 plastic model.
\tref{tab:ep_parameters} summarizes the material properties of the constituents.
\fref{fig:alloy} shows the realization of the alloy and a bundle of \ADD{representative} stress-strain responses to the cyclic loading \eqref{eq:F_samples}.
The hysteresis displays minimal hardening and rate dependence.
\cref{jones2024multiscale} has more details of this RVE.

\begin{table}[h]
\centering
\begin{tabular}{|l|c|}
\hline
Property & value \\
\hline
\hline
\textbf{Silicon} & \\
\quad Young's modulus     & 140  [GPa]\\
\quad Poisson's ratio    &   0.3  \\
\quad volume fraction    &   8.3\% \\
\hline
\textbf{Aluminum} & \\
\quad Young's modulus     & 70 [GPa]\\
\quad Poisson's ratio    & 0.32  \\
\quad Yield stress       & 120 [MPa] \\
\quad Hardening modulus  & 10.0 [MPa]  \\
\quad Exponential hardening coefficient   & 0.27 \\
\hline
time step          & 0.005 [s] \\
\hline
\end{tabular}
\caption{Elastic-plastic parameters}
\label{tab:ep_parameters}
\end{table}

\subsection{Glass bead reinforced viscoelastic silicone (VE)}
\begin{figure}
\centering
\begin{subfigure}[c]{0.35\linewidth}
\includegraphics[width=0.95\linewidth]{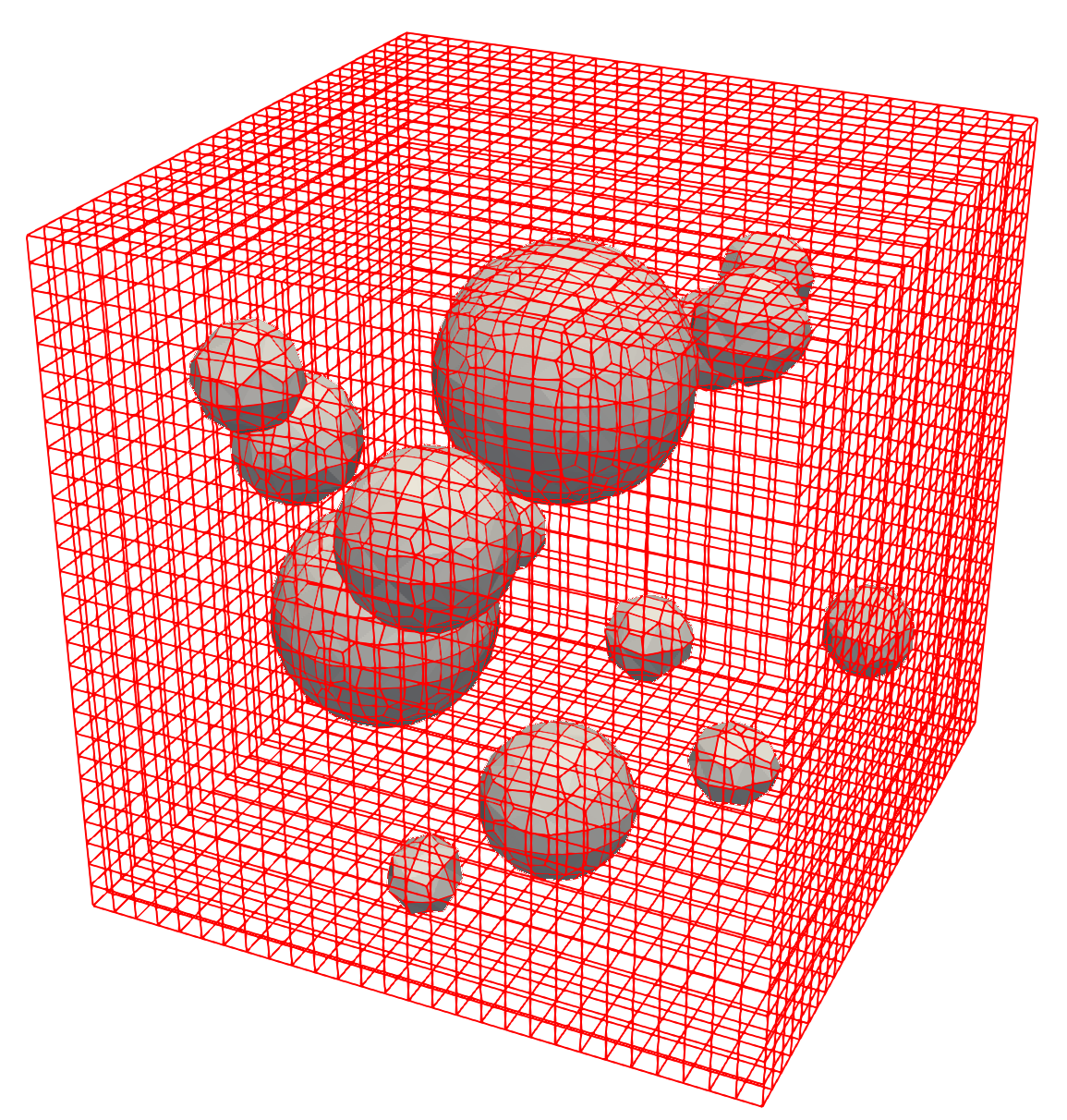}
\end{subfigure}
\begin{subfigure}[c]{0.5\linewidth}
\includegraphics[width=0.95\linewidth]{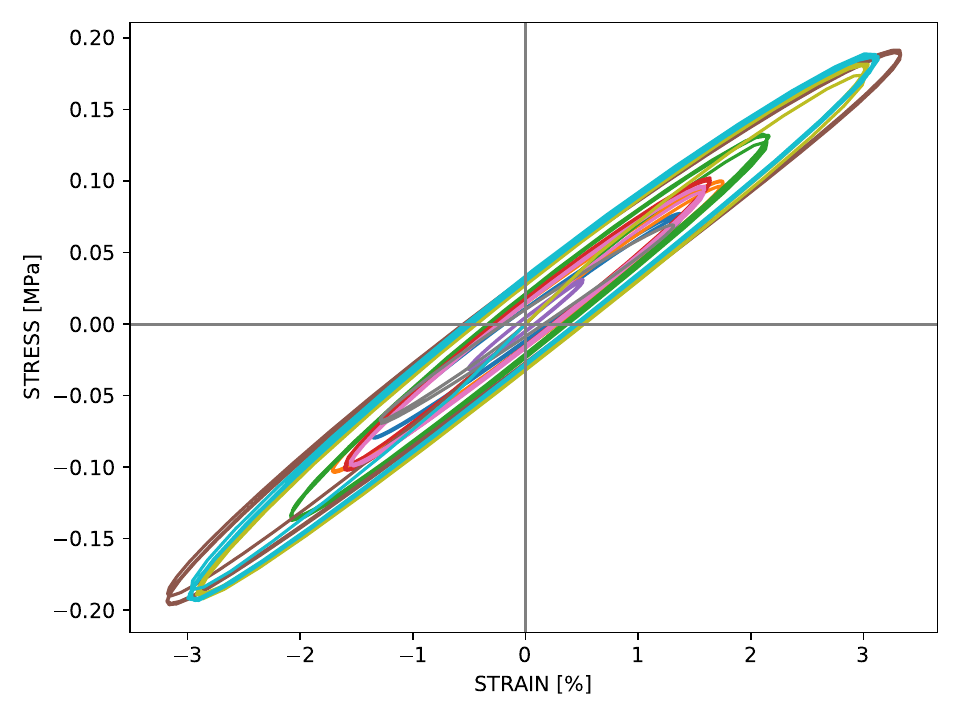}
\end{subfigure}
\caption{Silicone-glass bead composite. \ADD{ (a) representative  volume element showing spherical glass inclusions, (b) representative hysteresis curves of a random ensemble where realizations are distinguished by color. } }
\label{fig:composite}
\end{figure}

Reinforced polymers are common engineering materials for applications that need flexible yet durable materials.
Composite of elastic glass of various radii and viscoelastic polymer, as can be seen in \fref{fig:composite}.
\tref{tab:ve_parameters} summarizes the material properties of the constituents.
\fref{fig:composite} shows the realization of the composite on an unstructured mesh with 25800 elements.
A St.Venant model was used for the glass, while the  \emph{Universal Polymer Model} (UPM) \cite{long2017linear}, a complex Prony series/discrete memory kernel model, was used for the polymer matrix.
The UPM model includes a wide variety of time-scales, and the hysteresis displays rate-dependence consistent with this viscoelastic behavior.
The hysteresis shown in \fref{fig:composite} is characteristic of a viscoelastic material; however, the elastic beads make the composite response non-rheologically simple.
\cref{jones2022neural} has more details.

\begin{table}[h]
\centering
\begin{tabular}{|l|c|}
\hline
Property & value \\
\hline
\hline
\textbf{Glass} & \\
\quad Young's modulus    & 60 [GPa]  \\
\quad Poisson's ratio    & 0.33  \\
\quad volume fraction    & 13.5\% \\
\hline
\textbf{Silicone} & \\
\quad bulk modulus      & 920 [MPa] \\
\quad glassy shear modulus   &   3.62 [MPa] \\
\quad rubbery shear modulus  &   0.84 [MPa] \\
\quad relaxation times  &  1 [$\mu$s] : 3.16 [ks] \\
\hline
time step    & 0.005 [s]\\
\hline
\end{tabular}
\caption{Viscoelastic composite parameters}
\label{tab:ve_parameters}
\end{table}

\subsection{Fe viscoplastic polycrystal (VP)}
\begin{figure}
\centering
\begin{subfigure}[c]{0.35\linewidth}
\includegraphics[width=0.95\linewidth]{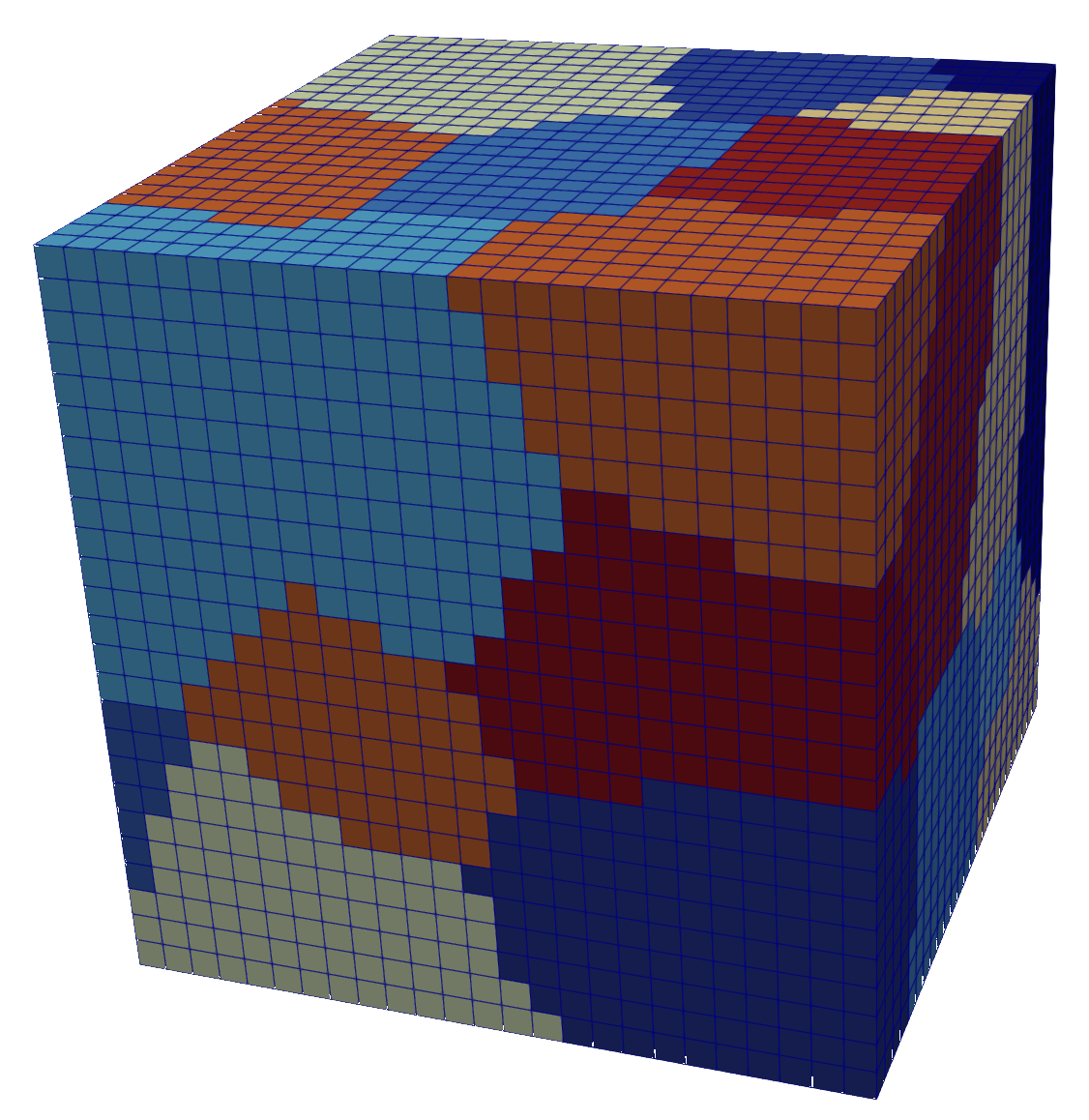}
\end{subfigure}
\begin{subfigure}[c]{0.5\linewidth}
\includegraphics[width=0.95\linewidth]{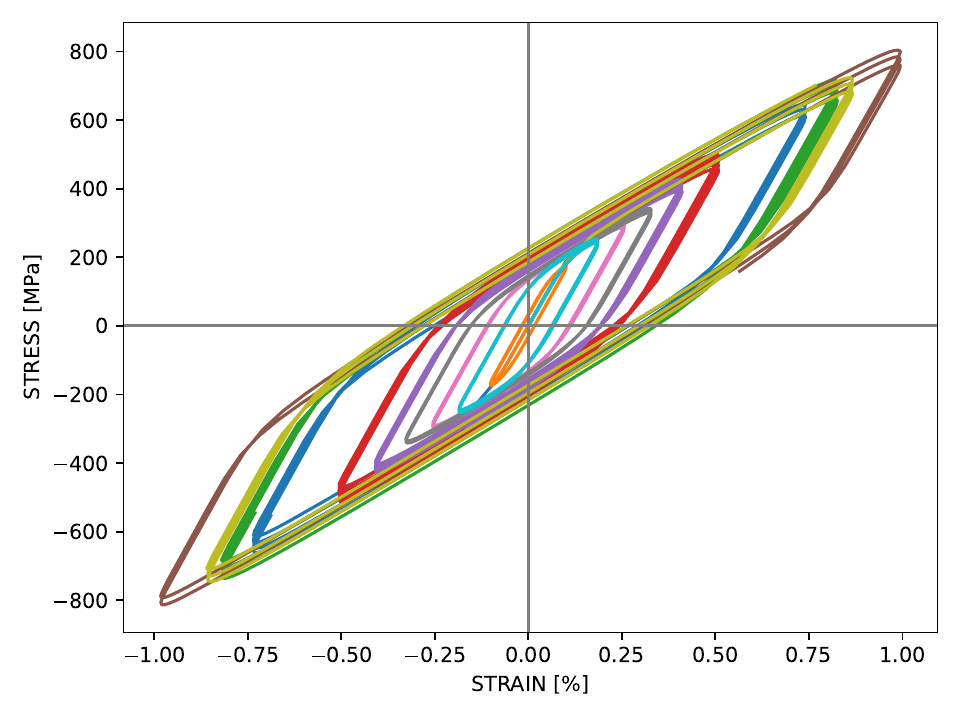}
\end{subfigure}
\caption{FCC Fe polycrystal.  \ADD{ (a) representative volume element showing polycrystalline structure colored by grain orientation, (b) representative hysteresis curves of a random ensemble where realizations are distinguished by color. } }
\label{fig:polycrystal}
\end{figure}

Crystal plasticity (CP) RVEs are often used as subgrid models of complex plastic behavior in metals.
We used a standard CP finite element \cite{roters2010crystal}  on a $25^3$ structured mesh to simulate a Fe (FCC) polycrystal.
A power-law flow rule and the well-known Kocks-Mecking hardening minus recovery \ADD{rule} \cite{mecking1981kinetics} were employed on all 12 slip planes.
\tref{tab:cp_parameters} summarizes the material properties.
\fref{fig:polycrystal} shows the realization of the polycrystal and the response curves where rate-dependence and plasticity are apparent.
\cref{jones2022neural} has more details.

\begin{table}[h]
\centering
\begin{tabular}{|l|r|}
\hline
Property & value \\
\hline
\hline
Elastic moduli & \\
\quad C11 & 204.6 [GPa] \\
\quad C12 & 137.7 [GPa] \\
\quad C44 & 126.2 [GPa] \\
\hline
Hardening modulus & 600 [MPa] \\
Recovery  modulus &   1 [MPa] \\
Initial hardening & 150 [MPa] \\
Rate exponent & 20 \\
\hline
time step    & 0.05 [s]\\
\hline
\end{tabular}
\caption{Crystal plasticity parameters for FCC Fe.}
\label{tab:cp_parameters}
\end{table}

\section{Results} \label{sec:results}

We used as identical as possible potentials for the DP, GSM, and MP NN implementations, where the differences are primarily due to the number of inputs to the dissipation potential prescribed by the frameworks, \ADD{ see \tref{tab:implementations}. }
\tref{tab:nn_specifics} summarizes the chosen architectural details.
The dimension of the hidden ISV space is the most arbitrary; the size was chosen based on previous studies \cite{jones2022neural,jones2025attention,jones2025physics}.
\ADD{We had to make specific selections for the architectural parameters, but these were guided preliminary studies so as not to unduly hamper any of the models in terms of under- or over-parameterization.}
\ADD{In preliminary and previous parameter exploration \cite{jones2022neural}, we found that the accuracy of the models is fairly insensitive to the size of the internal state vector, as it seems that redundancies are handled robustly by correlated state evolution.}
\ADD{In fact, the free energy potentials for each model are identical, and the number of parameters in the dissipation potentials are within 3\% of each other.}
\ADD{The following results should be viewed in light of these specifics.}

\ADD{We used explicit/forward Euler to integrate the flow rules in order to simplify the computational graph in training.
Predictions with other integrators are typically acceptable, for example using implicit/backward Euler, midpoint or Heun incurs only 1.86-2.88\% additional error over the VE validation data.}
Furthermore, we used the ADAM optimizer \cite{kingma2014adam} with a learning rate of 0.001 for 80,000 epochs with an early-stopping patience of 8,000 epochs.
Predictions were made with the weights that achieved the best in-training test losses.
The datasets were split into 80\% training data, 10\% in-training test data, and 10\% post-training validation data.
All results depicted in this section are for predictions on the held-out validation data \ADD{visualizations are for models with median validation scores}.
\ADD{Reported errors are root means squared (RMSE) errors relative to the same norm of the data, which is also normalized for training.}

\begin{table}[h]
\centering
\begin{tabular}{|l|c|ccc|}
\hline
potential & $\freeenergy$ & & $\dissipationpotential$ & \\
&               & DP & GSM & MP \\
\hline
\hline
number of ISVs       & 6   &    & 6 &   \\
input dimension      & 9   & 18 & 9 & 12 \\
number of layers     & 2   &    & 2 & \\
number of parameters & 345 & 1606 & 1545 & 1550 \\
activation           & softplus & & softplus & \\
\hline
\end{tabular}
\caption{NN potential parameters. Input dimensions depend on formulation hence the number of parameters in the DP, GSM, and MP implementations.}
\label{tab:nn_specifics}
\end{table}

\subsection{Elastoplasticity (EP) RVE data}
\begin{figure}
\centering
\begin{subfigure}[c]{0.99\linewidth} \centering
\includegraphics[width=0.99\linewidth]{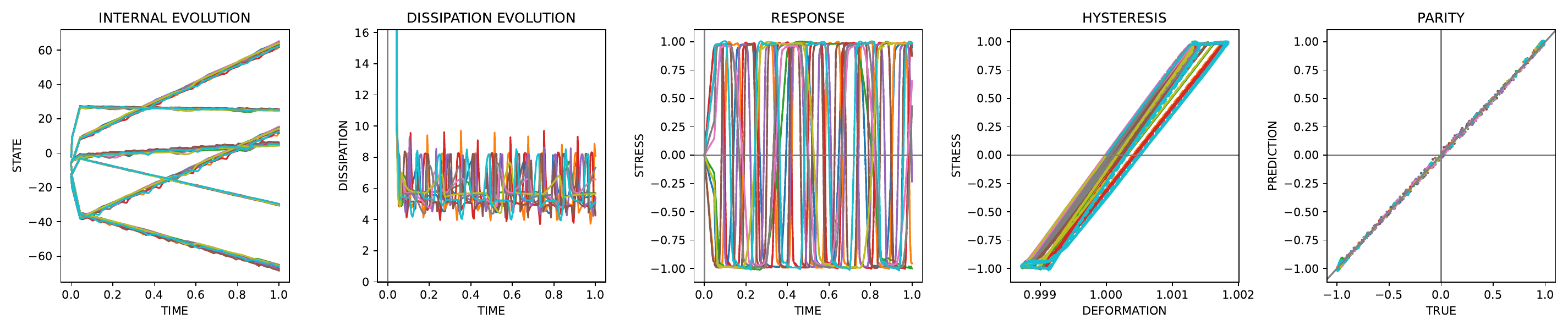}
\caption{DP}
\end{subfigure}
\begin{subfigure}[c]{0.99\linewidth} \centering
\includegraphics[width=0.99\linewidth]{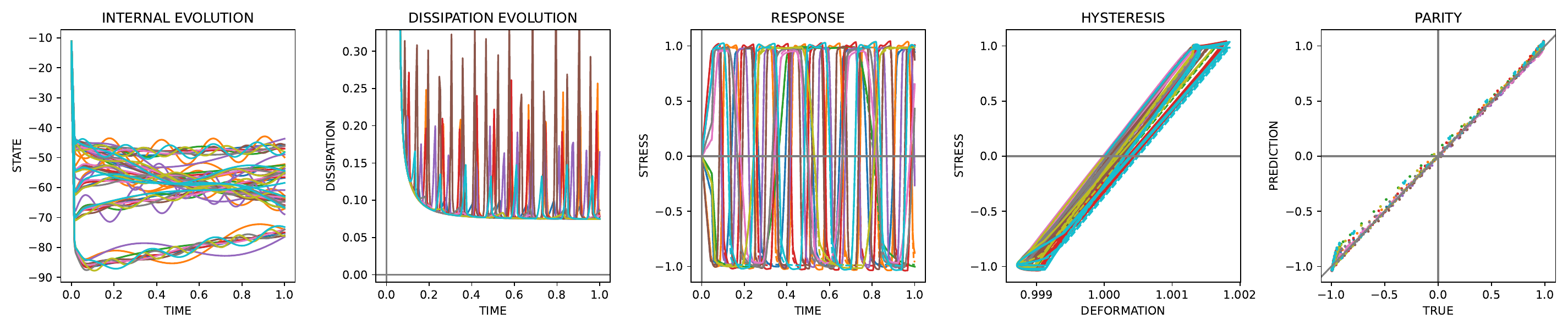}
\caption{GSM}
\end{subfigure}
\begin{subfigure}[c]{0.99\linewidth} \centering
\includegraphics[width=0.99\linewidth]{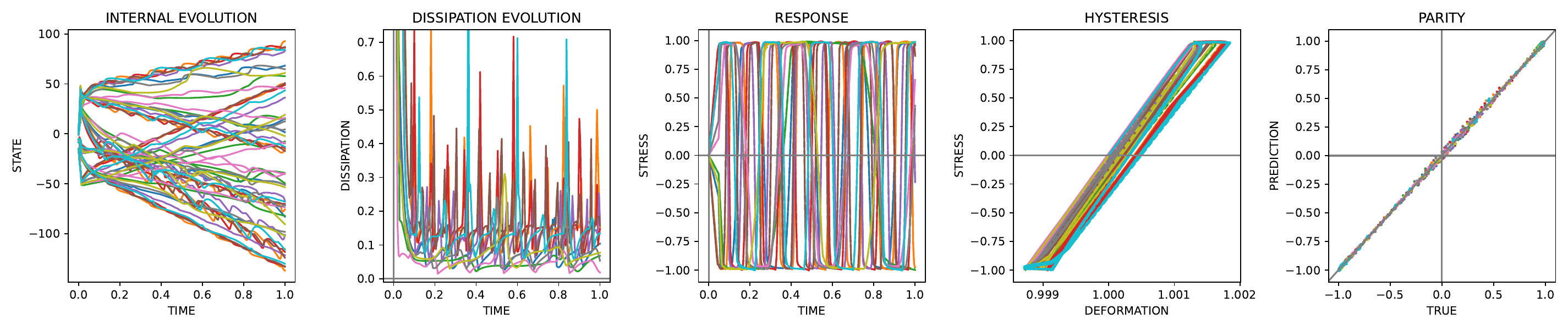}
\caption{MP}
\end{subfigure}
\caption{EP prediction comparison. \ADD{A representative randomly-selected subset of the validation data is shown for clarity, where realizations are distinguished by color; the same subset is used for each method. }}
\label{fig:EP}
\end{figure}

\fref{fig:EP} shows representative predictions on held-out EP validation data for each of the model frameworks.
This is arguably the most challenging phenomenon to represent due to the switching between conservative/elastic and dissipative/plastic behavior.
All three models predict the validation data well.
The ability of GSM models to represent elasto-plasticity apparently depends on the ability of the dissipation potential to represent gradient jumps at the origin \cite{mcbride2018dissipation} and the resulting subdifferential.
In preliminary studies, we tried a \emph{ReLU} activation for the dissipation potential, but it did not perform as well as the \emph{Softplus} we ultimately used.
The relaxed potential formulation of Holthusen and Kuhl \cite{holthusen2026complement} may have utility with this issue.

The dissipation predicted from each model shows high-frequency content commensurate with the loading; however, the magnitudes are significantly different for each of the models, and the dissipation does not go to zero during any phase.
\ADD{As observed in \cref{jones2025attention}, in general this indicates that the flow model has trouble representing no-flow/elastic states perfectly.}
This erroneous feature can be fixed with gating of the flow \cite{jones2025attention}  \ADD{and/or a duality penalty \cite{jones2025physics}}; however, no cause for the disagreement in the dissipation response between models, such as different units, is obvious.
The strong linear trends in the evolution of the internal variables for the predictions may be the intermediate cause of the dissipation offset.
Clearly, each model is discovering hidden states with different evolution patterns, \ADD{although, generally speaking, each is a harmonic variation around a transient trend followed by a linear phase.}

\subsection{Viscoelasticity (VE) RVE data}
\begin{figure}
\centering
\begin{subfigure}[c]{0.99\linewidth} \centering
\includegraphics[width=0.99\linewidth]{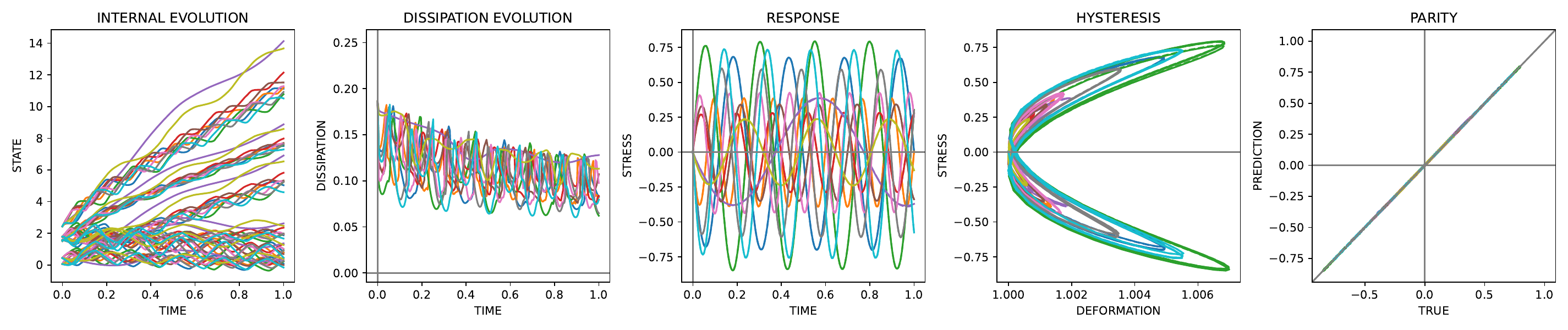}
\caption{DP}
\end{subfigure}
\begin{subfigure}[c]{0.99\linewidth} \centering
\includegraphics[width=0.99\linewidth]{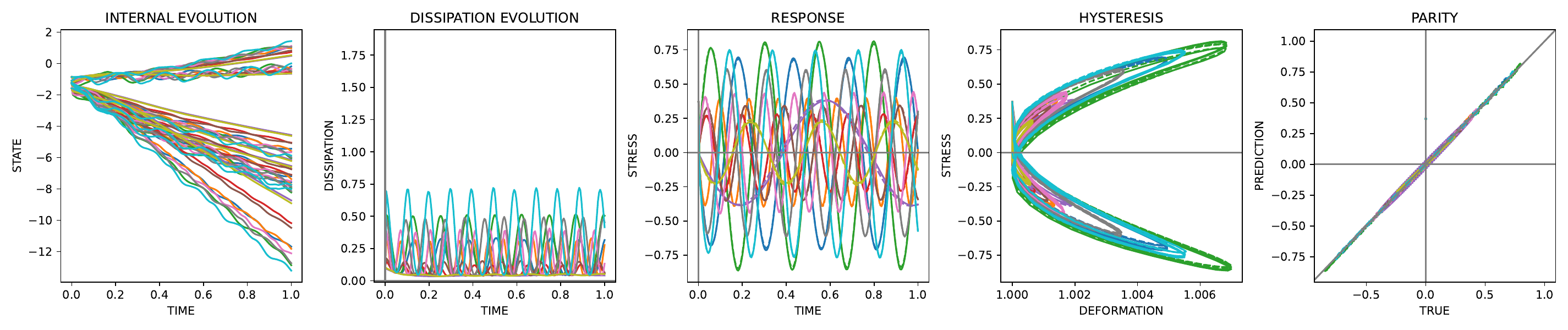}
\caption{GSM}
\end{subfigure}
\begin{subfigure}[c]{0.99\linewidth} \centering
\includegraphics[width=0.99\linewidth]{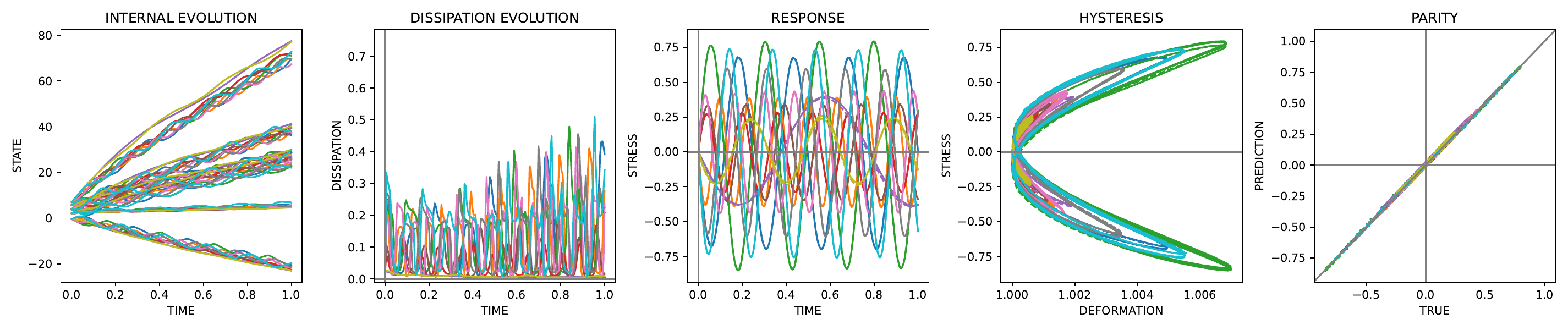}
\caption{MP}
\end{subfigure}
\caption{VE prediction comparison.  \ADD{A representative randomly-selected subset of the validation data is shown for clarity, where realizations are distinguished by color; the same subset is used for each method. }}
\label{fig:VE}
\end{figure}

\fref{fig:VE} shows predictions of held-out VE validation data.
All predictions are nearly perfect in accuracy, which is expected since the ODE-like framework of each model is closest to that of classical viscoelasticity.
Again, we see differences in the dissipation, with DP showing a slowly decreasing trend which conforms to physical expectations, GS is nearly constant in mean after a transient, and MP shows dissipation growth in some of the trials.
Like in the EP example, here all models have strong linear trends with superposed oscillations in their internal state trajectories.
Notably, DP has some internal state trajectories with steady means.
We conjecture that the strong linear trends may be evidence that only a lower-dimensional manifold in internal state space is needed to represent the phenomenology and, hence, certain aspects of the evolution are not constrained by data.

\subsection{Viscoplasticity (VP) RVE data}
\begin{figure}
\centering
\begin{subfigure}[c]{0.99\linewidth} \centering
\includegraphics[width=0.99\linewidth]{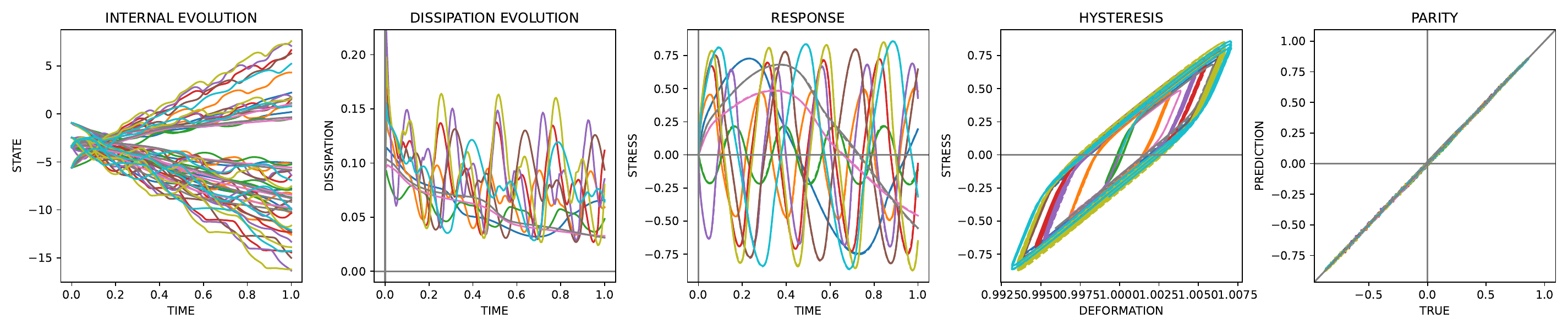}
\caption{DP}
\end{subfigure}
\begin{subfigure}[c]{0.99\linewidth} \centering
\includegraphics[width=0.99\linewidth]{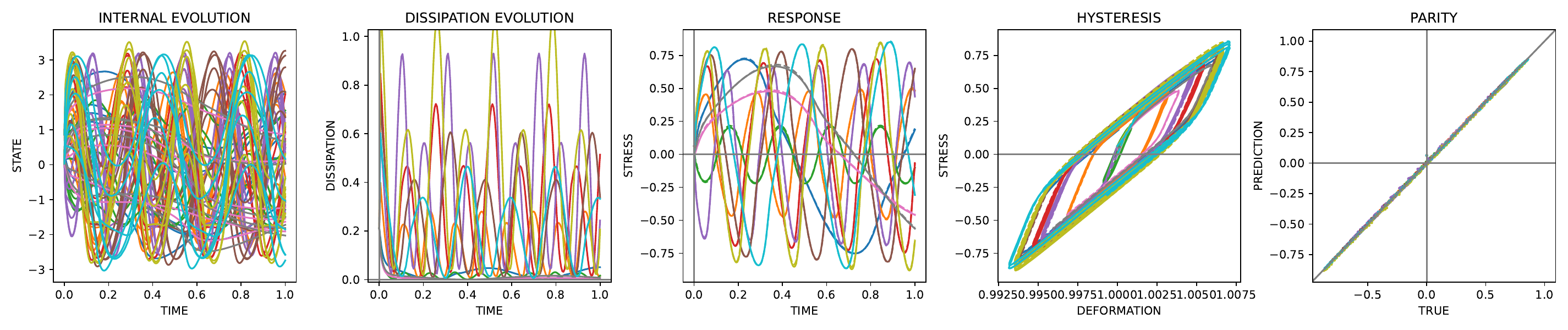}
\caption{GSM}
\end{subfigure}
\begin{subfigure}[c]{0.99\linewidth} \centering
\includegraphics[width=0.99\linewidth]{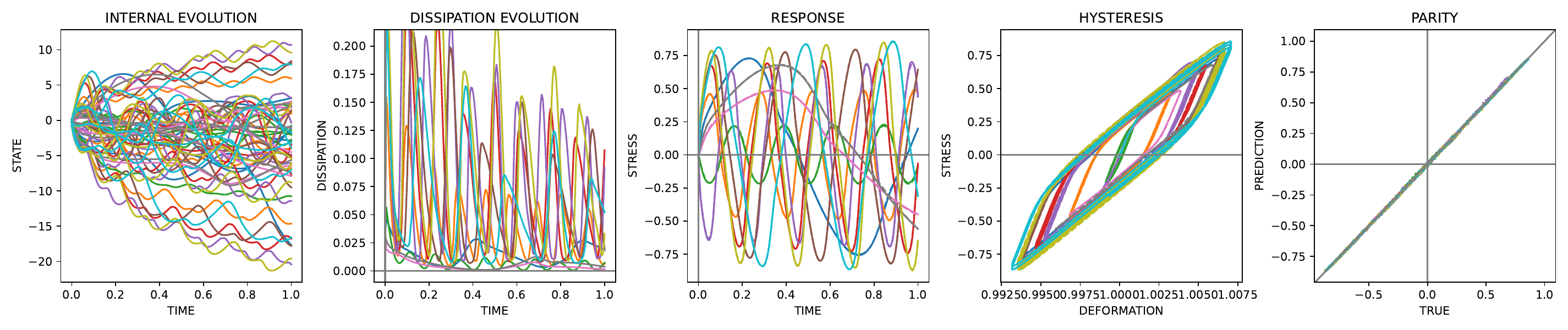}
\caption{MP}
\end{subfigure}
\caption{VP prediction comparison.  \ADD{A representative randomly-selected subset of the validation data is shown for clarity, where realizations are distinguished by color; the same subset is used for each method. }}
\label{fig:VP}
\end{figure}

\fref{fig:VP} shows representative predictions on held-out VP validation data for the VP dataset, where, like the VE data, the model predictions are highly accurate.
Also we, \ADD{again, observe} differences in the dissipation trajectories across the three models, although in this case, there is a decreasing trend to the MP dissipation, while DP and GS resemble their VE responses.
Interestingly, the internal state trajectories for each of the models are qualitatively different: DP has strong linear trends, GS has nearly steady oscillations, while MP has some trajectories with mild growth or decay.
These results seem to corroborate the conjecture that, depending on data and the presumed size of the internal state space, there are ambiguities in the dynamics due to the sensitivities of the observable stress to these states.

\section{Discussion} \label{sec:discussion}

\begin{figure}
\centering
\begin{subfigure}[c]{0.99\linewidth} \centering
\includegraphics[width=0.80\linewidth]{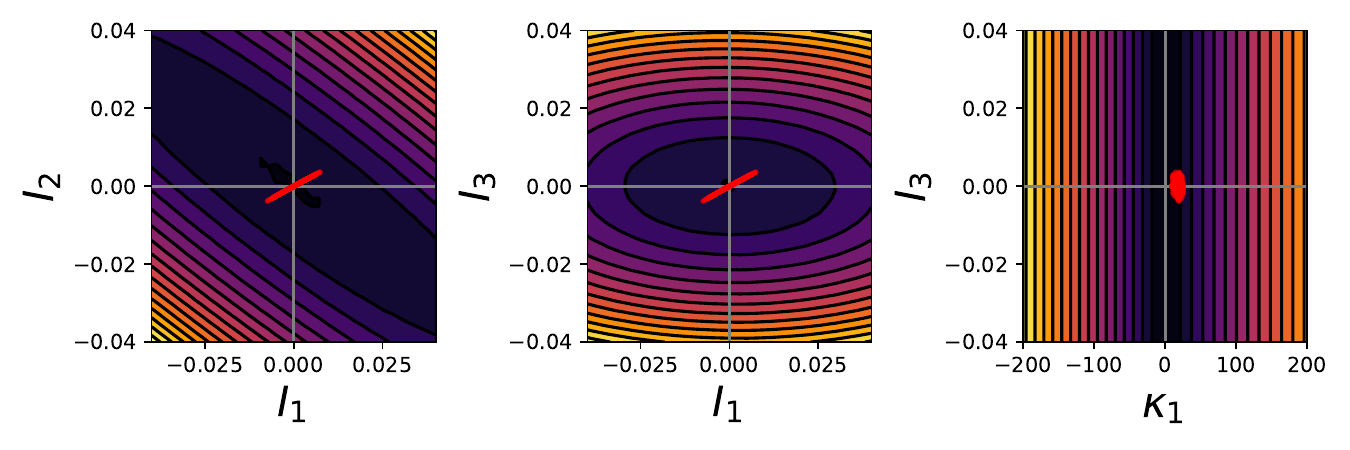}
\caption{DP}
\end{subfigure}
\begin{subfigure}[c]{0.99\linewidth} \centering
\includegraphics[width=0.80\linewidth]{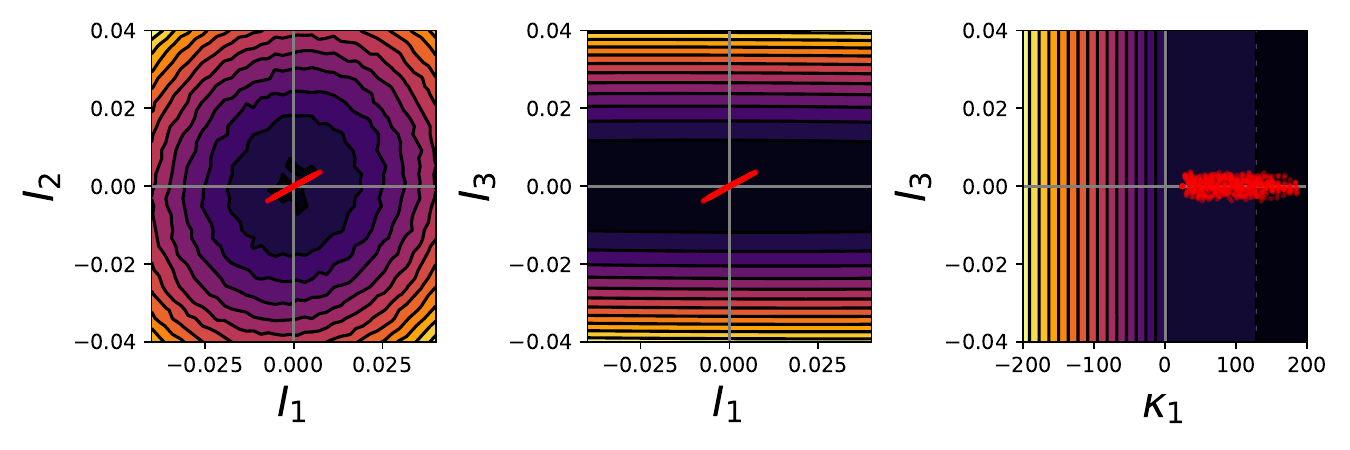}
\caption{GSM}
\end{subfigure}
\begin{subfigure}[c]{0.99\linewidth} \centering
\includegraphics[width=0.80\linewidth]{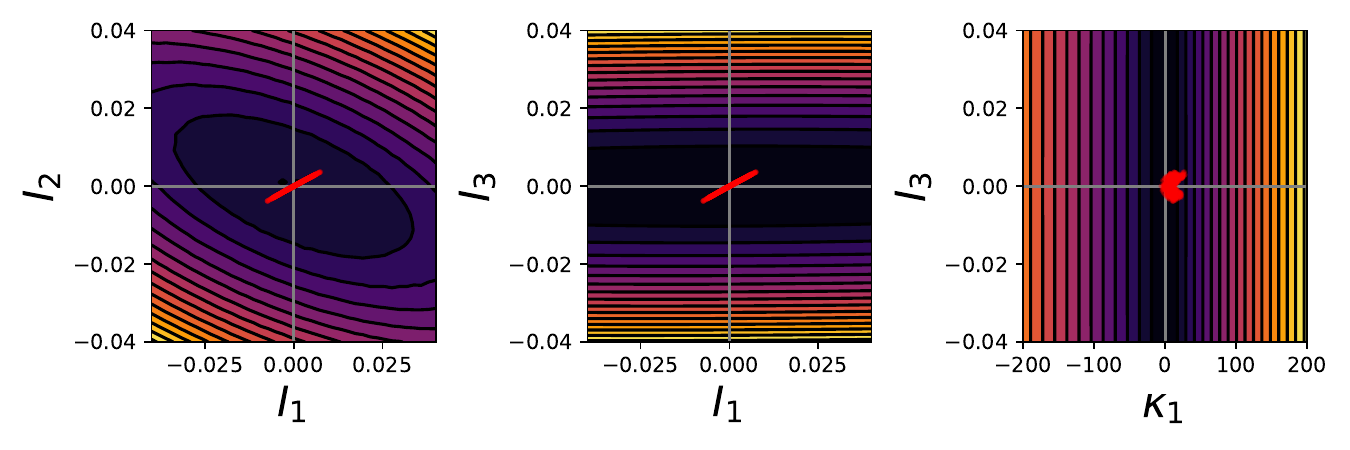}
\caption{MP}
\end{subfigure}
\caption{\ADD{Representative qualitative diagnostic of the learned} free energy $\freeenergy$ \ADD{surface. The contour plots are projected at the reference state \eqref{eq:ref_state}, while the training samples traverse the entire invariant--internal-state space; therefore, the visualization should not be interpreted as reproducible identification of the full potential surface. Note all invariants have been shifted to have their reference at zero for comparison.}  }

\label{fig:psi_viz}
\end{figure}

To continue with the investigation of the commonalities and distinctions between the three formulations, we examine (projections) of the potentials.
\fref{fig:psi_viz} shows the learned free energy for DP, GSM, and MP, taking \ADD{a median training to} the VP dataset as representative.
Recall that each framework shares the same NN representation for $\freeenergy$, although, in GSM, the $\dissipationpotential$ also contributes to the stress.
The particular potentials all resemble each other, especially given the fact that the limited locus of the accessible invariants depicted by the red samples \cite{fuhg2022physics}; i.e., gradients along this locus match across the potentials.
\ADD{These visualizations should be interpreted as particular qualitative diagnostics rather than reproducible identification of the full potential surface, especially given the fact that individual internal state trajectories are not uniquely determined.}
The main differences are not in the observable invariant projections, but in the internal state aspects, which necessarily differ given the different dynamics.
\ADD{Although there are clear correlations between the internal state dynamics of the models and within the individual states of a given model, the fact that they are distinct indicates they may be playing different roles in the various models.}
We also note that due to the fungibility of the internal state components, the internal state component of one model may be most highly correlated with a different component of another; this complicates comparison of the relatively high dimensional internal states.

\begin{figure}
\centering
\includegraphics[width=0.5\linewidth]{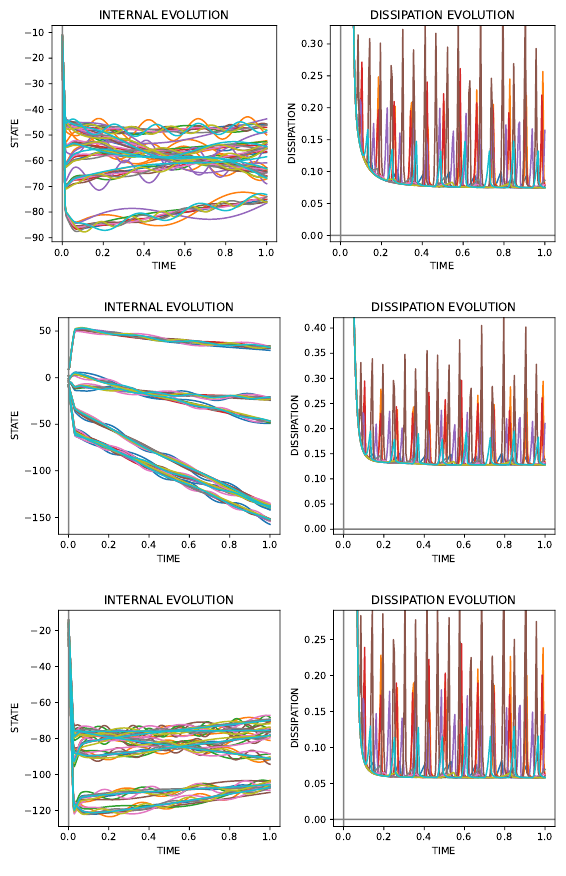}
\caption{\ADD{Variance in internal state evolution for GS model trained to EP data.}}
\label{fig:dissipation}
\end{figure}

\ADD{
On this point, even different trainings of the same model will exhibit the fungibility, and potentially the overparameterization, of the internal states as \fref{fig:dissipation} illustrates for the GS model.
The hidden state trajectories only resemble each other in that they tend to group and have quasi-linear phases.
Furthermore, although the dissipation for these three model instantiations resemble each other in that we observe something like exponential decay to a steady state, even the steady state varies slightly from instantiation to instantiation.
We attribute this to dissipation being a unobserved, derived quantity, hence, the addition of the dissipation to the training objective would refine the model calibration but require the measurement of generated heat and related dissipative mechanisms.
}

\begin{figure}
\centering
\begin{subfigure}[c]{0.99\linewidth} \centering
\includegraphics[width=0.80\linewidth]{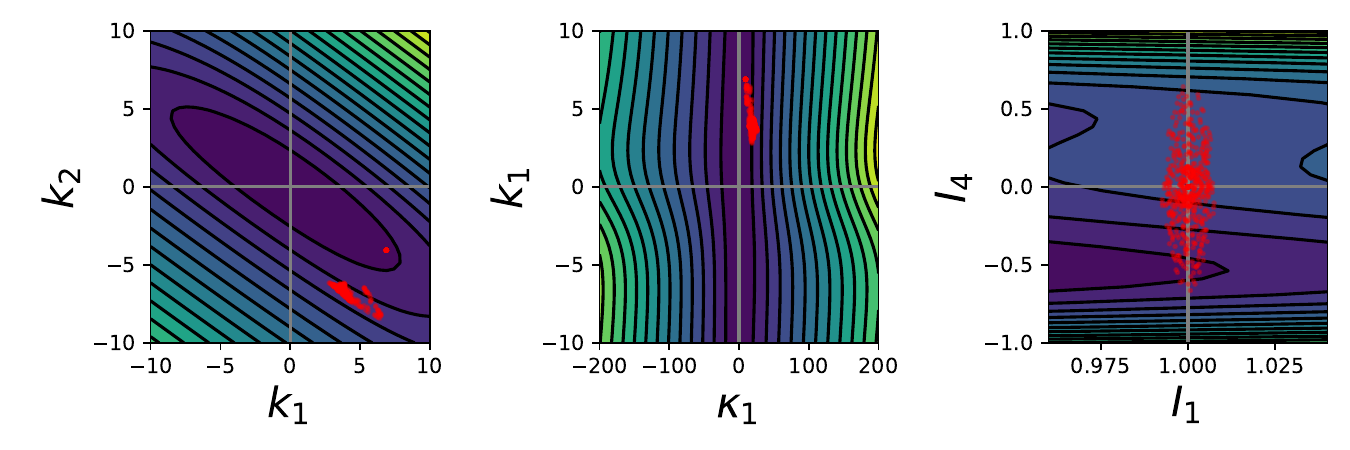}
\caption{DP}
\end{subfigure}
\begin{subfigure}[c]{0.99\linewidth} \centering
\includegraphics[width=0.80\linewidth]{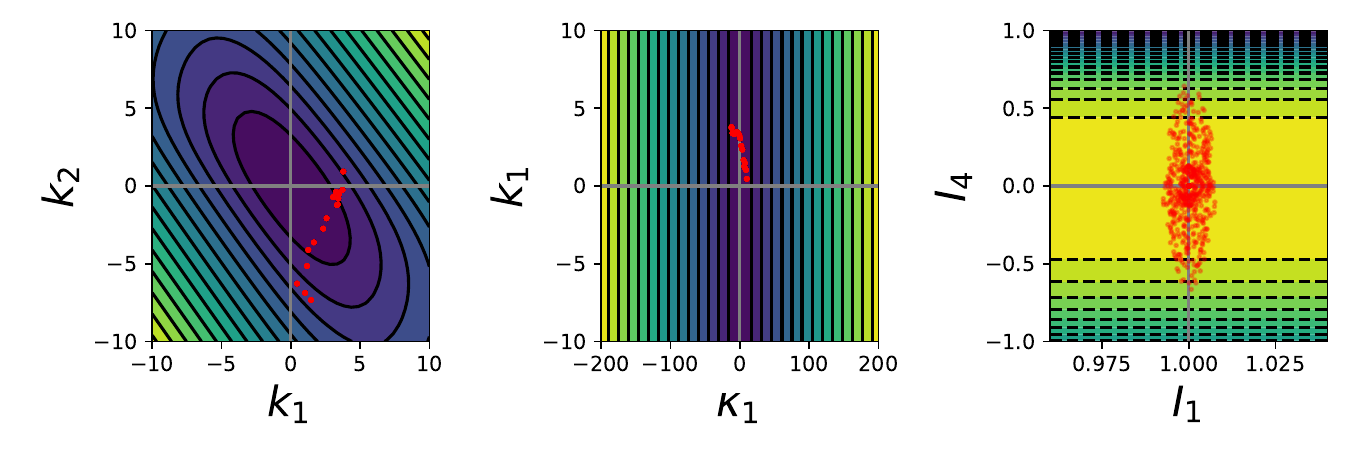}
\caption{GSM}
\end{subfigure}
\begin{subfigure}[c]{0.99\linewidth} \centering
\includegraphics[width=0.80\linewidth]{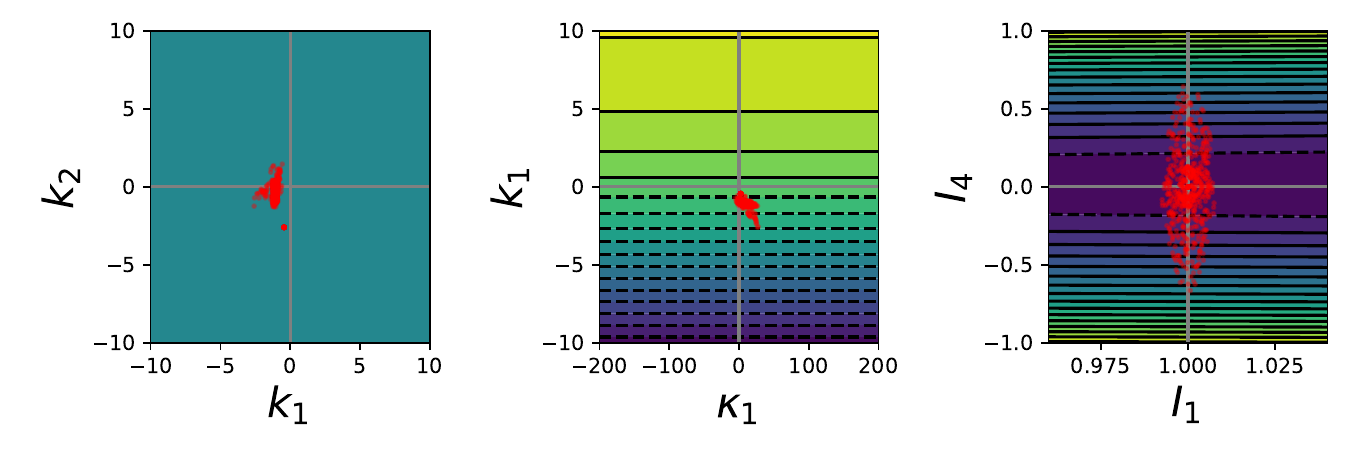}
\caption{MP}
\end{subfigure}
\caption{
\ADD{Representative qualitative diagnostic of the learned} dissipation potential $\dissipationpotential$ \ADD{surface, with contour plots projected at the reference state \eqref{eq:ref_state} while the training samples traverse the entire invariant--internal-state space. The visualization reflects locally constrained behavior near the sampled locus rather than identification of the full high-dimensional potential.}
}
\label{fig:phi_viz}
\end{figure}

As expected \ADD{from the differences summarized in \tref{tab:formulation_distinctions} and \ref{tab:implementations}}, the fitted dissipation potentials shown in \fref{fig:phi_viz} show more qualitative differences than the free energy.
The left panel shows the projection of $\dissipationpotential$ onto the first two components of the conjugate force $\conjugateforce$, while the middle panel shows the projection onto the first component of the internal state $\internalvariables$ and conjugate force.
The right panel shows a projection onto the observable deformation $I_1$ and rate $I_4$ invariants.
The dissipation potentials in the observable projection largely resemble each other \ADD{despite the concavity requirement for the GSM $\dissipationpotential$}, with the strongest sensitivities in rate invariant $I_4$.
The DP shows some deviation from convexity in $I_4$, \ADD{unlike the MP $\dissipationpotential$ and the enforced concavity in the GSM model}.
The projections in the other planes are radically different, in part due to the fungibility of the internal state components and the corresponding conjugate forces.
Nevertheless, the flatness of the MP in the 1-2 conjugate force plane is remarkable, which seems to indicate it is not using these components to represent the data.
In contrast, the enforced convexity of the DP and GSM in this plane is apparent.
Of some interest is the variation of the DP dissipation potential in the $\internalvariables_1$-$\conjugateforce_1$ projection.
Overall, it appears that the NNs are learning fairly simple, nearly constant/linear/quadratic dependencies \ADD{and the behavior away from the training data is due to the embedded physical constraints.}

\begin{figure}
\centering
\begin{subfigure}[c]{0.99\linewidth} \centering
\includegraphics[width=0.99\linewidth]{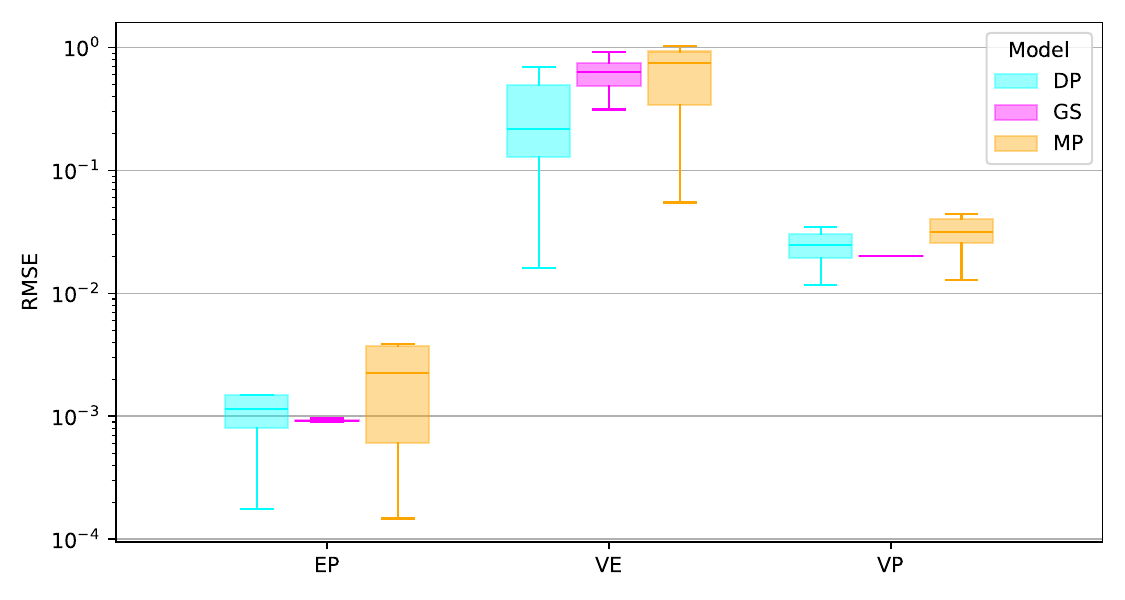}
\caption{RVE/no closed form}
\end{subfigure}
\begin{subfigure}[c]{0.99\linewidth} \centering
\includegraphics[width=0.99\linewidth]{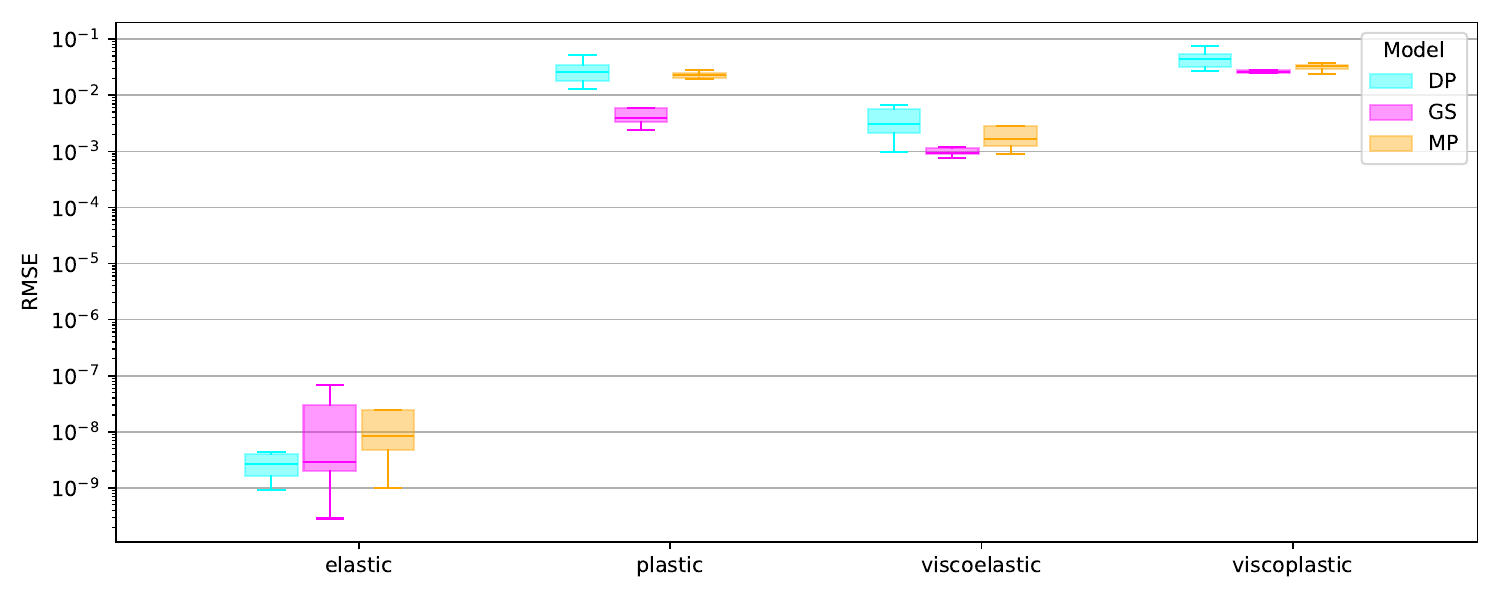}
\caption{Homogeneous/closed form}
\end{subfigure}
\caption{Summary of validation errors for 10 independent trainings.}
\label{fig:summary}
\end{figure}

We now make a statistical assessment of the performance of each model on held-out data.
\fref{fig:summary}a summarizes the 3 formulations (DP, GS, MP) performance across the 3 RVE-based datasets described in \sref{sec:data} for multiple training sessions where the train/test/validation data has been shuffled, as have the initial weights.
Generally, all models give good representations and predictions, although the GSM seems to underperform relative to the least constrained DP formulation across the three datasets we used.
To investigate whether this ranking is due to the lack of a closed form / simple dependence of the data, we ran the same study on 4 benchmark datasets: (a) hyperelastic  Neohookean basis, which was the basis for the other 3 models, (b) J2 finite elastoplasticity with linear hardening, (c)  memory kernel-type viscoelasticity, and (d) Perzyna-type viscoplastic model based on (b).
Each data-generating model used a standard implementation \cite{simo1998computational}, and more details can be found in \cref{jones2025attention}.
Interestingly \fref{fig:summary}b shows the trend, loosely speaking, reverses with GSM moderately outperforming the other two formulations.
It is easy to conjecture that the more constrained structure of the GSM framework is responsible for these findings, i.e., it represents clean stress-strain history relations well but is mildly hampered if the data does not satisfy some of the strict embedded requirements, such as associative flow \cite{lemaitre1994mechanics,point2013convex}.
\ADD{In fact, others have indicated that certain requirements, such as polyconvexity, may also be too restrictive for RVE data \cite{klein2026limitations,wollner2026concurrent}.}

\section{Conclusion} \label{sec:conclusion}

We presented a unified thermodynamic and machine learning comparison of three constitutive frameworks: dissipation potential (DP), generalized standard materials (GSM), and metriplectic (MP).
By embedding each framework within \ADD{common neural potential architecture, we reduced architectural and training mismatch and focused the comparison on the influence of thermodynamic structure.}
The theoretical and numerical comparison revealed a hierarchy of structural restriction.
The DP formulation enforces convexity in the conjugate force and guarantees dissipation while allowing flexible state dependence.
GSM additionally imposes Legendre--Fenchel duality and normality, restricting admissible evolution through convexity in both forces and rates.
MP introduces operator-based reversible and irreversible flows that separate energy-conserving and entropy-producing dynamics.
Across the elastoplastic, viscoelastic, and viscoplastic RVE datasets,  theory-based structural bias proved to create accurate models, although in some cases it limited expressive flexibility.
GSM provided the most tightly constrained and thermodynamically prescribed formulation, while DP achieved comparable predictive accuracy with fewer structural assumptions.
Our interpretation of MP offered a geometrically consistent evolution framework and naturally accommodates operator-based dynamics, and gave predictions on par with the other representations while generating different internal dynamics.
These results demonstrate that thermodynamic model selection acts as an inductive bias in data-driven constitutive modeling that aids generalization behavior.
\ADD{Although, generally speaking, we believe the overall PANN recipe for material model construction, i.e. :
\begin{itemize}
\item identify physical constraints,
\item select a thermodynamic framework which guides the information flow,
\item select appropriate inputs, invariants, and outputs,
\item design function representations for the framework components that respect the constraints and are complete, or at least expressive, function classes,
\item let the remaining parametric information be data driven through calibration
\end{itemize}
is more important than the particular framework choice and brings material modeling to the level of a methodological science.}

In future work, we will investigate learnable operators in the MP setting, implicit standard material (ISM) formulations via bipotentials, and extensions to nonlocal \cite{junker2021extended} and inverse problems.
This is in pursuit of the larger goal of a unified and extensible machine learning framework that can accommodate the spectrum of solid material phenomenology, which can distinguish between rate-dependent and rate-independent inelastic responses and identify the anisotropic structure required by the data, while preserving thermodynamic admissibility by construction.
The MP operator-based decomposition cleanly separates reversible and irreversible mechanisms and suggests a pathway toward learning generalized skew and metric operators that remain thermodynamically admissible by construction.%
\footnote{
The MP formulation provides a structured setting for the evolution of tensorial internal variables.
For example, for a symmetric second-order tensor $\Ab$, the reversible contribution may take the form
\begin{equation*}
\dot{\Ab} = \Omegab \Ab - \Ab \Omegab, \qquad  \Omegab = \dot{\Rb} \Rb^T,
\end{equation*}
which corresponds to the action of a skew operator on the space of second-order tensors and represents an objective, energy-conserving evolution.
Here $\Rb$ is the local rotation from the polar decomposition of the deformation gradient.
}
The fact that ISM can represent DP, GSM, and MP formulations indicates it has the promise to be a more encompassing formulation than the formalisms we implemented, albeit with some numerical challenges to surmount.
Both avenues have the potential to be superior data-driven PANNs for material response.

\section*{Acknowledgements}
\indent
We would like to thank Nat Trask~(UPenn) and Anthony Gruber~(Sandia) for insights into metriplectic learning methods.
\ADD{We also thank the two anonymous reviewers who provided constructive critiques and pointed out shortcomings in our original presentation.}

\indent
This work was funded by the Laboratory Directed Research and Development (LDRD) program at Sandia National Laboratories; this funding is gratefully acknowledged.
Sandia National Laboratories is a multi-mission laboratory managed and operated by National Technology \& Engineering Solutions of Sandia, LLC (NTESS), a wholly owned subsidiary of Honeywell International Inc., for the U.S. Department of Energy’s National Nuclear Security Administration (DOE/NNSA) under contract DE-NA0003525. This written work is authored by an employee of NTESS. The employee, not NTESS, owns the right, title and interest in and to the written work and is responsible for its contents. Any subjective views or opinions that might be expressed in the written work do not necessarily represent the views of the U.S. Government. The publisher acknowledges that the U.S. Government retains a non-exclusive, paid-up, irrevocable, world-wide license to publish or reproduce the published form of this written work or allow others to do so, for U.S. Government purposes. The DOE will provide public access to results of federally sponsored research in accordance with the DOE Public Access Plan.

\indent
This material is based upon work partially supported by the U.S. National Science Foundation under award No. 2452029.
The opinions, findings, and conclusions, or recommendations expressed are those of the authors and do not necessarily reflect the views of the NSF.
The authors acknowledge the Texas Advanced Computing Center (TACC) at The University of Texas at Austin for providing computational resources that have contributed to the research results reported within this paper.

\bibliographystyle{unsrt}


\appendix
\numberwithin{equation}{section}

\section{Implicit standard material (ISM) theory}\label{appendix:ISM}
The ISV framework leaves the evolution map $\flow$ largely unrestricted beyond the dissipation requirement.
Implicit standard materials \cite{de1992generalisation,bodoville2001implicit,de2002implicit} introduce additional structure by replacing the explicit flow rule with an implicit force--flux relation generated by a \emph{bipotential}.

\paragraph{Definition (Bipotential)}
A function
\begin{equation}
\bipotential = \bipotential(\conjugateforce,\dot{\internalvariables}; \Zc)
\end{equation}
is called a bipotential if, in addition to lower semicontinuity, it satisfies the following properties:
\begin{enumerate}

\item \textbf{Separate convexity}
of the implicit functions:
\begin{align}
\dot{\internalvariables} \mapsto \bipotential(\conjugateforce,\dot{\internalvariables})
&\quad \text{is convex for fixed } \conjugateforce, \\
\conjugateforce \mapsto \bipotential(\conjugateforce,\dot{\internalvariables})
&\quad \text{is convex for fixed } \dot{\internalvariables}.
\end{align}

\item \textbf{Generalized Fenchel inequality:}
\begin{equation}
\bipotential(\conjugateforce, \dot{\internalvariables})
\;\geq\;
\conjugateforce \cdot \dot{\internalvariables}
\quad
\text{for all } (\conjugateforce, \dot{\internalvariables}) \ .
\end{equation}

\end{enumerate}
Here, $\Zc= \{\lagrangestrain, \internalvariables, \abstemperature, \ldots\}$ denotes a collection of auxiliary variables, which we suppress in the remainder of this section.
Geometrically, the bipotential defines a surface in the $(\dot{\internalvariables},\conjugateforce)$ space that lies above the bilinear form $\conjugateforce \cdot \dot{\internalvariables}$.
The admissible evolution corresponds to those points where the two surfaces coincide.
The conjugate force remains determined by the free energy as in the DP formulation,
\begin{equation}
\conjugateforce = -\partial_{\internalvariables} \freeenergy \ .
\end{equation}
The role of the bipotential is to define the admissible evolution of the internal variables through an implicit force--flux relation.
Given the thermodynamic force $\conjugateforce$, the evolution rate $\dot{\internalvariables}$ is required to satisfy
\begin{equation}
\dot{\internalvariables} \in \partial_{\conjugateforce} \bipotential(\conjugateforce,\dot{\internalvariables}).
\end{equation}
Equivalently, admissible pairs $(\conjugateforce,\dot{\internalvariables})$ lie on the constitutive graph characterized by
\begin{equation}
\conjugateforce \in \partial_{\dot{\internalvariables}} \bipotential,
\qquad
\dot{\internalvariables} \in \partial_{\conjugateforce} \bipotential .
\end{equation}
Thermodynamic constraints then follow immediately from the generalized Fenchel inequality, since
\begin{equation}
\dissipation
=
\,\conjugateforce \cdot \dot{\internalvariables}
=
\,\bipotential(\conjugateforce, \dot{\internalvariables})
\;\ge\; 0.
\end{equation}
If the bipotential is therefore normalized such that
\begin{equation}
\bipotential(\conjugateforce, \dot{\internalvariables})
\geq 0,
\end{equation}
on admissible pairs, thermodynamic consistency is ensured.

We now examine specializations of ISM that correspond to common frameworks.

\paragraph{Dissipation potential (DP)}
In the classical dual dissipation potential formulation, the force--flux relation is generated by a single convex potential depending only on the conjugate force. \ADD{To express this relation in bipotential form, let $\dissipationpotential^*$ denote the Legendre--Fenchel conjugate of $\dissipationpotential$,
\begin{equation}
\dissipationpotential^*(\dot{\internalvariables})
=
\sup_{\conjugateforce}
\left(
\conjugateforce \cdot \dot{\internalvariables}
-
\dissipationpotential(\conjugateforce)
\right).
\end{equation}
We then write the corresponding separable bipotential as
\begin{equation}
\beta(\conjugateforce, \dot{\internalvariables})
=
\dissipationpotential(\conjugateforce)
+
\dissipationpotential^*(\dot{\internalvariables}) .
\end{equation}}
The dual dissipation potential is assumed to satisfy:
\begin{equation}
\dissipationpotential(\conjugateforce)
\ \text{is convex, lower semicontinuous},
\qquad
\dissipationpotential(\zerob)=0 \ .
\end{equation}
Admissible evolution is characterized by the subdifferential relation
\begin{equation}
\dot{\internalvariables}
\in
\partial_{\conjugateforce} \dissipationpotential(\conjugateforce).
\end{equation}
Equivalently,
\begin{equation}
\conjugateforce
\in
\partial_{\dot{\internalvariables}}
\dissipationpotential^*(\dot{\internalvariables}).
\end{equation}
By convexity of $\dissipationpotential$, the Fenchel inequality yields
\begin{equation}
\dissipationpotential(\conjugateforce)
+
\dissipationpotential^*(\dot{\internalvariables})
\ge
\conjugateforce \cdot \dot{\internalvariables}.
\end{equation}
Equality holds precisely on admissible force--flux pairs. Thus, on the constitutive graph,
\begin{equation}
\beta(\conjugateforce,\dot{\internalvariables})
=
\conjugateforce\cdot\dot{\internalvariables}.
\end{equation}
Since $\dissipationpotential(\zerob)=0$ and $\dissipationpotential$ is assumed non-negative and minimized at $\zerob$, its conjugate is also non-negative under the corresponding normalization. Consequently, on admissible pairs,
\begin{equation}
\dissipation
=
\conjugateforce \cdot \dot{\internalvariables}
=
\beta(\conjugateforce,\dot{\internalvariables})
\ge 0 \ .
\end{equation}
The associated Fenchel gap,
\begin{equation}
\dissipationpotential(\conjugateforce)
+
\dissipationpotential^*(\dot{\internalvariables})
-
\conjugateforce\cdot\dot{\internalvariables}
\ge 0,
\end{equation}
vanishes when
$\dot{\internalvariables}\in\partial_{\conjugateforce}\dissipationpotential(\conjugateforce)$.

\paragraph{Generalized standard material (GSM)}
The classical associative GSM structure is recovered when the bipotential is chosen in separable Fenchel form, i.e.,
\begin{equation}
\beta(\conjugateforce,\dot{\internalvariables}) =
\dissipationpotential^*(\dot{\internalvariables}) + \dissipationpotential(\conjugateforce),
\end{equation}
where $\dissipationpotential^*$ and $\dissipationpotential$ form a Legendre--Fenchel dual pair. In particular, \ADD{the Legendre--Fenchel conjugate $\dissipationpotential$ of $\dissipationpotential^*$ is defined by}
\begin{equation}
\dissipationpotential(\conjugateforce)
=
\sup_{\dot{\internalvariables}}
\left( \conjugateforce \cdot \dot{\internalvariables} - \dissipationpotential^*(\dot{\internalvariables}) \right) \ ,
\end{equation}
and conversely
\begin{equation}
\dissipationpotential^*(\dot{\internalvariables}) =
\sup_{\conjugateforce} \left( \conjugateforce \cdot \dot{\internalvariables} - \dissipationpotential(\conjugateforce) \right) \ ,
\end{equation}
By construction, this duality implies the Fenchel inequality
\begin{equation}
\dissipationpotential^*(\dot{\internalvariables}) + \dissipationpotential(\conjugateforce) \ge
\conjugateforce \cdot \dot{\internalvariables}.
\end{equation}
Equality holds if and only if $\conjugateforce$ and $\dot{\internalvariables}$ are dual variables, i.e.,
\begin{equation}
\conjugateforce \in \partial_{\dot{\internalvariables}} \dissipationpotential^{*}(\dot{\internalvariables})
\quad \Longleftrightarrow \quad
\dot{\internalvariables} \in \partial_{\conjugateforce} \dissipationpotential(\conjugateforce)
\end{equation}
which yields associative normality.
Under this condition, dissipation reduces to
\begin{equation}
\dissipation = \conjugateforce \cdot \dot{\internalvariables} \ge 0 \ .
\end{equation}

This duality is not available for a general bipotential.
A general bipotential $\beta(\conjugateforce,\dot{\internalvariables})$ is only assumed to be separately convex and to satisfy the generalized Fenchel inequality
\begin{equation}
\beta(\conjugateforce, \dot{\internalvariables}) \ge \conjugateforce \cdot \dot{\internalvariables}.
\end{equation}
In contrast to the separable GSM choice $\beta=\dissipationpotential+\dissipationpotential^*$, an arbitrary (non-separable) $\beta$ does not define a convex function of $\dot{\internalvariables}$ whose Legendre--Fenchel transform produces a dual convex function of $\conjugateforce$.
Equivalently, for general $\beta$ there need not exist potentials $\dissipationpotential^*(\dot{\internalvariables})$ and $\dissipationpotential(\conjugateforce)$ such that
\begin{equation}
\beta(\conjugateforce,\dot{\internalvariables})
=
\dissipationpotential^*(\dot{\internalvariables})+\dissipationpotential(\conjugateforce) \ ,
\end{equation}
which is tied to the identity \eqref{eq:dissipation_dual_identity}.
The separable Legendre–Fenchel form is therefore a special subclass of the general bipotential structure. In GSM, this primal–dual structure is a constitutive requirement, whereas in the DP formulation above it provides an equivalent bipotential representation of the prescribed force–flux relation

\paragraph{Metriplectic form}

A metriplectic-type metric relation can be represented by a quadratic bipotential for the irreversible metric component. We consider
\begin{equation}
\beta(\conjugateforce,\dot{\internalvariables})
=
\conjugateforce\cdot\dot{\internalvariables}
+
\frac{1}{2}
\left(
\dot{\internalvariables}
-
\Mb\,\conjugateforce
\right)
\cdot
\Mb^{-1}
\left(
\dot{\internalvariables}
-
\Mb\,\conjugateforce
\right),
\end{equation}
where the metric operator $\Mb = \Mb(\lagrangestrain,\internalvariables)$ is symmetric positive semi-definite:
\begin{equation} \label{eq:pd}
\Mb = \Mb^T \succ 0 \ .
\end{equation}
Admissibility enforces the metric relation
\begin{equation}
\dot{\internalvariables} = \Mb\,\conjugateforce.
\end{equation}
The resulting dissipation becomes
\begin{equation}
\dissipation
= \conjugateforce\,\cdot \,\dot{\internalvariables}
= \conjugateforce\, \cdot \Mb \, \conjugateforce
\ge 0 \ ,
\end{equation}
which follows from the positive definiteness of the $\Mb$ operator \eqref{eq:pd}.

\end{document}